\documentclass[12pt]{article}
\usepackage[ascii]{inputenc}
\usepackage{amsmath,amssymb,amsfonts,amsthm}
\usepackage[margin=2.8cm]{geometry}
\usepackage{graphicx}
\usepackage{bbm}
\usepackage[caption=false]{subfig}
\usepackage[pdftex,bookmarks=false,colorlinks=true,linkcolor=blue,
citecolor=blue,filecolor=black,urlcolor=blue]{hyperref}

\providecommand{\llangle}{\langle\!\langle}
\providecommand{\rrangle}{\rangle\!\rangle}

\numberwithin{equation}{section}

\begin{document}

\title{Current fluctuations for anharmonic chains\\ in thermal equilibrium}

\author{Christian B. Mendl and Herbert Spohn
\footnote{Zentrum Mathematik and Physik Department,
Technische Universit\"at M\"unchen,
Boltzmannstra{\ss}e 3, 85747 Garching bei M\"unchen, Germany
Email: \href{mailto:mendl@ma.tum.de}{mendl@ma.tum.de}, \href{mailto:spohn@ma.tum.de}{spohn@ma.tum.de}}}

\date{December 8, 2014}

\maketitle

\begin{abstract}
\vspace{1cm}We study the total current correlations for anharmonic chains in thermal equilibrium, putting forward predictions based on the second moment sum rule and on nonlinear fluctuating hydrodynamics. We compare with molecular dynamics simulations for hard collision models. For the first time we investigate the full statistics of time-integrated currents. Generically such a quantity has Gaussian statistics on a scale $\sqrt{t}$. But if the time integration has its endpoint at a moving sound peak, then the fluctuations are suppressed and only of order $t^{1/3}$. The statistics is governed by the Baik-Rains distribution, known already from the fluctuating Burgers equation.
\end{abstract}

\newpage
\section{Introduction}\label{sec1}

One-dimensional classical fluids have anomalous transport properties, as has been recognized already in the early 1970ies \cite{AlWa1970, PoRe75, Ernst1976}. But intense theoretical investigations, very much supported by molecular dynamics (MD), started in the late 1990ies \cite{LeLi1997}. To study energy transport, a popular setup is to impose thermal boundary conditions. If the boundary temperatures differ, a steady state energy flux is enforced. Fourier's law states that the energy flux is proportional to $1/L$ with $L$ the system size. The anomaly consist in having a larger energy flux, namely $\sim L^{-1+\tilde{\alpha}}$ with $\tilde{\alpha} >0$. A related property is the time-dependent response in the local energy to a small perturbation close to the origin of the equilibrated system, which can be expressed through the energy-energy time-correlation in equilibrium. Denoting the microscopic fluctuating energy density by $e(x,t)$, the said correlation is $\langle e(x,t)e(0,0)\rangle - \langle e(0,0)\rangle\langle e(0,0)\rangle= S_{33}(x,t)$, where the average is over the canonical equilibrium state and stationarity in $x,t$ has been used. $S_{33}$ has a central heat peak together with left and right sound peaks propagating with the speed of sound, $c$. Fourier's law means a diffusive, $\sqrt{t}$, spreading of the heat peak, while the anomaly corresponds to a super-diffusive spreading. For large times the heat peak is expected to maintain the scaling form $(\lambda_{\mathrm{h}}t )^{-\gamma} f_{\mathrm{h}}( (\lambda_{\mathrm{h}}t )^{-\gamma}x)$ with some scaling exponent $\gamma> \frac{1}{2}$ and a non-Gaussian scaling function $f_{\mathrm{h}}$. $\lambda_{\mathrm{h}}$ is a model-dependent scale coefficient. The structure of time-correlations for the conserved fields has been discussed in \cite{Spo, MeSp13} with MD support in \cite{MS14}. Here we will concentrate on current fluctuations, a topic which has hardly been touched upon in the mentioned investigations.

The local energy satisfies the conservation law 
\begin{equation}\label{1.1}
\partial_t e(x,t) + \partial_x j_e(x,t) = 0\,,
\end{equation}
where $ j_e(x,t)$ is the microscopic fluctuating energy current. Using stationarity in $x$ and in $t$, one concludes that 
\begin{equation}\label{1.2}
\partial_t ^2\langle e(x,t) e(0,0)\rangle = \partial_x^2 \langle j_e(x,t) j_e(0,0)\rangle\,.
\end{equation}
Thus, in essence, the local energy correlation carries the same information as the local current correlation. However in the Green-Kubo formula the total fluctuating current,
\begin{equation}\label{1.3}
j_{e,L}(t) = \frac{1}{\sqrt{L}}\int _0^L dx\, \big(j_e(x,t) - \langle j_e(0,0)\rangle \big)\,,
\end{equation}
is employed, more precisely its correlation $\langle j_{e,L}(t)j_{e,L}(0)\rangle$. Now the connection to $S_{33}$ is less direct and a separate investigation is required. One-dimensional fluids have three conservation laws and, in principle, one should study the entire $3\times3$ correlator of total currents.

Generally it is assumed that, even for anomalous transport, the steady state energy flux enforced through thermal boundary conditions is still related to the total equilibrium energy current correlation as 
\begin{equation}\label{1.3a}
\int_0^{L/c} dt \langle j_{e,L}(t)j_{e,L}(0)\rangle\,.
\end{equation}
Thus along with a direct MD measurement of the energy flux, mostly also the energy current correlation in thermal equilibrium is simulated. The contributions are numerous, too many to be listed here, and we refer to the reviews \cite{LeLi2003, Dhar2008}. Generically one tries to estimate the long-time decay of the correlation, respectively the small frequency behavior of its Fourier transform. However the full $3\times 3 $ correlation matrix has never been studied, with the notable exception \cite{vBPo} who discuss also the total momentum current correlation. In our contribution we discuss the full correlation matrix and compare the predictions based on fluctuating hydrodynamics with MD simulation of the three hard collision models from \cite{MS14}.

The Green-Kubo type formula \eqref{1.3a} uses only the variance of $j_{e,L}(t)$. Since $j_{e}(x,t)$ is a fluctuating quantity, also its distribution is of physical interest. An example would be the distribution of the time-integrated current,
\begin{equation}\label{1.4}
\int_0^t ds\, j_e(x,s)\,,
\end{equation}
at some given spatial location $x$. For transport in electronic systems such current statistics are of great physical interest \cite{LeLev1995}. For stochastic dynamics with a single conservation law similar quantities have been studied in considerable detail. Results on typical fluctuations \cite{FeSp05} and large deviations \cite{Der06} are available. For mechanical systems, to our knowledge the only study concerns higher-order cumulants for a fluid with hard core particles and alternating masses \cite{DerridaGe2010}. Our novel ansatz is to consider the time-integrated current along the straight line $ s \mapsto (sv,s),\, 0 \leq s \leq t$. Generically such a current has Gaussian statistics on the scale $\sqrt{t}$. But if the endpoint is located at one of the peaks, i.e., $v = 0,\pm c$ with $c$ the speed of sound, and if one chooses the current corresponding to the respective peak, then the size of the fluctuations will be much smaller and the distribution becomes non-Gaussian.

To provide on outline, basic material is collected in Sect.~\ref{sec2}. The predictions based on mode-coupling and on the second moment sum rule are explained in Sect.~\ref{sec3}. They are compared with MD simulations of the hard collision models studied already in \cite{MS14}, see Sect.~\ref{sec4}. We study the full current statistics in Sects.~\ref{sec5} and \ref{sec6}. In particular, we explain the Baik-Rains distribution and provide numerical evidence for such a behavior.

\section{Current correlations, basic properties}\label{sec2}

Most of the material in this section can be found in earlier literature. But we could not find an exposition which, from our perspective, is sufficiently concise. At the risk of repetition, we list items for later use. 
We consider an anharmonic chain with interaction potential $V$. $V$ is bounded from below and satisfies the one-sided bound $V(x) \geq c_0 + c_1 \lvert x \rvert$ for either $x > 0$ or $x<0$ with some $c_1 >0$. Then, at fixed inverse temperature $\beta >0$, there is
a non-empty interval $I$ such that the one-particle partition function
\begin{equation}\label{2.1}
Z_1 = \int_{-\infty}^\infty dx\, e^{-\beta (V(x) + Px)} < \infty
\end{equation}
for $P\in I$. The dynamics of the chain is governed by the hamiltonian
\begin{equation}\label{2.2}
H = \sum_j \big( \tfrac{1}{2 m} p_j^2 + V(q_{j+1} - q_j)\big)\,.
\end{equation} 
Here $p_j$ is the momentum and $q_j$ the position of the $j$-th particle. We consider the case where all particles have the same mass $m$. The extension to alternating masses, i.e., to a unit cell containing two particles, will be discussed below. In principle, the difference $q_{j+1} - q_j$ could be bounded, or half-bounded, which would be encoded by letting 
$V (x) \to \infty$ as $x$ approaches a boundary point of such an interval. It is convenient to introduce the stretch $r_j = 
q_{j+1} - q_j$. Then the equations of motion read
\begin{equation}\label{2.3}
\frac{d}{dt}r_j(t) = \tfrac{1}{m} p_{j+1}(t) - \tfrac{1}{m} p_j(t)\,,\quad \frac{d}{dt}p_j(t) = V'(r_j(t)) - V'(r_{j-1}(t))\,.
\end{equation} 
 The local energy is defined by $e_j = \tfrac{1}{2m} p_j^2 + V(r_j)$ and satisfies 
\begin{equation}\label{2.4}
\frac{d}{dt}e_j(t) = \tfrac{1}{m} p_{j+1}(t)\, V'(r_j(t)) - \tfrac{1}{m} p_j(t)\, V'(r_{j-1}(t))\,.
\end{equation} 
Thus there are three local conservation laws and we require that there exist no further ones. Unfortunately this condition is extremely difficult to check. However it does rule out integrable chains as the nearest neighbor Toda and harmonic chain.

In essence there are two different definitions for what one means by an instantaneous local current. The first, seemingly more physical, prescription is based on interpreting $q_j$
as the position of a particle in the physical space $[0,L]$ with periodic boundary conditions. Then $j(x,t)$ is the current of 
the conserved field at location $x\in [0,L]$ at time $t$. It has two contributions, one is merely the flow of the conserved quantity across $x$, while the second one is slightly nonlocal and involves the force acting between two particles close to $x$.
For example the energy density reads
\begin{equation}\label{2.3a}
e(x,t) = \sum_{j=0}^{N-1} \delta(q_j-x) \big(\tfrac{1}{2m} p_j^2 + V(r_j) \big)\,.
\end{equation}
The energy current is then determined through \eqref{1.1} and is given by 
\begin{multline}\label{2.3b}
j_{e}(x,t) = \sum_{j=0}^{N-1} \delta(q_j-x)\, \tfrac{1}{m} p_j \big(\tfrac{1}{2m} p_j^2 + V(r_j)\big) \\ 
- \sum_{j=0}^{N-1} \tfrac{1}{m} p_{j+1}\, V'(r_j) \big(\chi(q_{j} \leq x \leq q_{j+1}) - \chi(q_{j+1} \leq x \leq q_j)\big)\,,
\end{multline}
where $\chi(\cdot) = 1$ if the condition in the argument is satisfied and $\chi(\cdot) = 0$ otherwise.
This is the Eulerian point of view. For a true one-dimensional fluid of unlabeled particles, the potential being the sum over all pairs $(i,j)$, there is no choice and the analogue of \eqref{2.3a}, \eqref{2.3b} is the appropriate definition. For anharmonic chains the particles are labeled and one can take also the Lagrangian point of view, in which $(r_j,p_j)$ is regarded as the ``field" at lattice site $j$. 
The currents then follow from the equations of motion, in the sense that the time change of a conserved field, when summed over the lattice interval $[j_ 1,j_2]$, is only through the currents at the two boundary points. In our example the energy current follows from Eq.~\eqref{2.4} and is given by
\begin{equation}\label{2.3c}
{\mathcal{J}}_3(j,t) = - \tfrac{1}{m} p_j(t)\, V'(r_{j-1}(t))\,.
\end{equation} 
There is no contribution from free streaming. Currents flow only if there is a force acting. As in \cite{Spo}, we will adopt here the lattice field theory point of view.

On a macroscopic level one expects the Lagrangian and Eulerian points of view to agree up to the obvious scale changes. For example, the asymptotic decay of a total current correlation should be the same in either definition. But the finite time corrections may differ. Also in MD simulations there will be finite size corrections which generically depend on the definition employed.

We consider the finite ring $[0,\dots,N-1] = \Lambda _N$ and impose periodic boundary conditions as $r_{N+j} = r_j$, $p_{N+j} = p_j$,
which is also the labeling used in the numerical part. For the infinite volume limit the centered ring $ \Lambda _N^\mathrm{c} = [-N,\dots,N]$ is more convenient. 
Initially the system is in canonical equilibrium. In MD simulations, the microcanonical measure is more natural and thus frequently used. We will discuss their relation below.
For canonical equilibrium
the intensive variables are the pressure, $P$, and the inverse temperature,
$\beta$. This state has the probability density 
\begin{equation}\label{2.6}
\big(\sqrt{2\pi\beta}\,Z_1\big)^{-N}\exp\!\Big[{-\beta} \sum_{j =0}^{N-1} \big( \tfrac{1}{2m} p_j^2 + V(r_j) + P r_j \big) \Big]
\prod_{j=0}^{N-1} dr_j\, dp_j\,.
\end{equation} 

The collection $\{r_j,p_j\}_{j = 0,\dots,N-1}$ are thus independent random variables with identical distribution. 
Of course, this carries over to the infinite volume limit. The condition \eqref{2.1} ensures that for $P \in I$ the partition 
function is finite. We have to introduce some notation for the averages. A simple $\langle \cdot \rangle$ means equilibrium average with the precise conditions specified through the context. To be more specific we use the subscripts $P,\beta$
and also indicate the finite volume $\Lambda_N$. Omitting $\Lambda_N$ refers to infinite volume $\mathbb{Z}$. Thus $\langle \cdot \rangle_{P,\beta}$ means the infinite volume canonical equilibrium average. To be systematic, one should also shift the $p_j$'s to the mean $\mathsf{u} \neq 0$, which however can be accounted for by a Galilean transformation.

From \eqref{2.3} and \eqref{2.4} one reads off the $3$-vector of conserved fields as
\begin{equation}\label{2.9a}
\vec{g}(j,t) = \big(r_j(t), p_j(t), e_j(t)\big)\,,
 \end{equation} 
 $ j \in \Lambda_N$, $ \vec{g} = (g_1,g_2,g_3)$, and the respective local currents as
\begin{equation}\label{2.9}
\vec{\mathcal{J}}(j,t) = - \Big( \tfrac{1}{m} p_j(t), V'(r_{j-1}(t)), \tfrac{1}{m} p_j(t) V'(r_{j-1}(t)) \Big) \,,
\end{equation}
 $ j \in \Lambda_N$, $ \vec{\mathcal{J}} = (\mathcal{J}_1,\mathcal{J}_2,\mathcal{J}_3)$. Currents and fields are related through the conservation law
\begin{equation}\label{2.10a}
\partial_t\,\vec{g}(j,t) + \vec{\mathcal{J}}(j+1,t)- \vec{\mathcal{J}}(j,t) = 0 \,, \mod\,\, \Lambda_N\,.
\end{equation}
Out of the vector in \eqref{2.9a} one forms the correlator of the conserved fields
\begin{equation}\label{2.9b}
S_{\alpha\alpha'}(j,t) = \big\langle g_\alpha(j,t); g_{\alpha'}(0,0) \big\rangle_{P,\beta} \,,
\end{equation}
using already stationarity in $j,t$, and correspondingly out of the vector in \eqref{2.9} the current-current correlator
\begin{equation}\label{2.9c}
\Gamma_{\alpha\alpha'}(j,t) = \big\langle \mathcal{J}_\alpha(j,t); \mathcal{J}_{\alpha'}(0,0) \big\rangle_{P,\beta} \,.
\end{equation}
Here $\alpha,\alpha' = 1,2,3$ is the label for the components  and $\langle X;Y \rangle = \langle XY\rangle - \langle X\rangle\langle Y\rangle$ denotes the second cumulant. These correlators have symmetries. They result from stationarity and invariance under time reversal together with the fact
that the fields 1,3 are even and the field 2 is odd, while current 2 is even and currents 1,3 are odd under the momentum reversal
$p_j \mapsto - p_j$ for all $j$. Therefore, $\mathrm{mod}\,\, \Lambda_N$,
\begin{equation}\label{2.9d}
\begin{split}
S_{\alpha\alpha'}(j,t) &= S_{\alpha'\alpha}(-j,-t)\,,\quad S_{\alpha\alpha'}(j,t) = (-1)^{\alpha+ \alpha'}S_{\alpha\alpha'}(j,-t) \\
\Gamma_{\alpha\alpha'}(j,t) &= \Gamma_{\alpha'\alpha}(-j,-t)\,,\quad \Gamma_{\alpha\alpha'}(j,t) = (-1)^{\alpha +\alpha'}\Gamma_{\alpha\alpha'}(j,-t)\,.
\end{split}
\end{equation}
It will be convenient to regard $S(j,t)$ and $\Gamma(j,t)$ as $3\times 3$ matrices.

The total currents are defined by
\begin{equation}\label{2.10}
\vec{\mathcal{J}}_{\mathrm{tot},\Lambda_N}(t) = \frac{1}{\sqrt{N}} \sum_{j=0}^{N-1} \vec{\mathcal{J}}(j,t ) 
\end{equation} 
and the total current covariance reads
\begin{equation}\label{2.11}
\Gamma_{\Lambda_N,\alpha\alpha'} (t) = \big\langle \mathcal{J}_{\mathrm{tot},\Lambda_N,\alpha}(t); \mathcal{J}_{\mathrm{tot},\Lambda_N,{\alpha'}}(0) \big\rangle_{\Lambda_N} = \sum_{j=0}^{N-1} \big\langle \mathcal{J}_\alpha(j,t); \mathcal{J}_{\alpha'}(0,0) \big\rangle_{\Lambda_N}\,.
\end{equation} 
$\mathcal{J}_1$ is itself conserved, hence $\mathcal{J}_{\mathrm{tot},1}(t)$ is independent of $t$ and
\begin{equation}\label{2.12}
\Gamma_{\Lambda_N,1\alpha}(t) = \big\langle \mathcal{J}_1(0,0); \mathcal{J}_\alpha(0,0) \big\rangle_{\Lambda_N}\,
\end{equation} 
for $\alpha =1,2,3$. The only total currents of interest are thus $\Gamma_{\Lambda_N,\alpha \alpha'} (t)$ with $\alpha, \alpha' = 2,3$.
From the symmetries in \eqref{2.9d} it follows that $\Gamma_{\Lambda_N,12} (t) =0$ and also 
$\Gamma_{\Lambda_N,23} (t) = -\Gamma_{\Lambda_N,23} (-t) = \Gamma_{\Lambda_N,32} (-t) = - \Gamma_{\Lambda_N,32} (t) $.

In our contribution we first take the limit $N\to\infty$ and then study the long time decay of the correlations.
The reversed point of view is considered in \cite{DerridaGe2010}. The two limits cannot be interchanged. For fixed $N$ and $t \to \infty$ the sound modes collide. In \cite{vBPo}, the dynamics is followed through 17 complete returns of the sound peaks to the origin. In this particular MD simulation the peaks seem to pass through each other with little interaction. Another option is to put both boundary points in contact with a Langevin type reservoir at the \emph{same} temperature \cite{DerridaGe2010}. Then the canonical measure, possibly with some boundary potentials, is the unique steady state. The sound mode is absorbed and partially reflected at the boundaries. This leads to an exponential damping of $\Gamma(t)$ as $e^{- t/\gamma}$ with $\gamma$ of the order $N/c$, $c$ the sound velocity, which masks the power law decay of the total current correlations.

We now consider the centered interval $\Lambda^{\mathrm{c}}_N$ and study the limit $N\to \infty$ in \eqref{2.11}. Up to an exponentially small error $\mathcal{J}_\alpha (j,t)$ should depend only on the variables in the cone $\{r_i, p_i : \lvert i-j\rvert < c_0\,t \mod N\}$ with some constant $c_0 > c$. Thus the current correlation on the right of \eqref{2.11} has an exponential decay uniform in $N$. Each term in the sum converges to its infinite volume limit. Hence by dominated convergence one concludes that
\begin{equation}\label{2.17}
\lim_{N \to \infty} \Gamma_{\Lambda^{\mathrm{c}}_N,\alpha\alpha'}(t) = \Gamma_{\alpha\alpha'}(t) = \sum_{j=-\infty}^\infty \big\langle \mathcal{J}_\alpha (j,t); \mathcal{J}_{\alpha'}(0,0) \big\rangle_{P,\beta} \,.
\end{equation}

\noindent\textit{Comment}. To our knowledge quasi-locality has been established only for anharmonic chains with a harmonic on-site potential \cite{Butta,Raz}. In our context it would be interesting to establish a similar result for interactions which depend only on positional differences.
\medskip

\noindent\textit{(i) sum rules}. Let $f: \mathbb{Z} \to \mathbb{R}$ be a test function and $\nabla_N f(j) = f(j+1) - f(j)$ with periodic boundary conditions relative to $\Lambda_N$. The time-integrated version of the conservation law \eqref{2.10a} then reads
\begin{equation}\label{2.10b}
\sum_{j \in \Lambda^{\mathrm{c}}_N}f(j)\big(g_\alpha(j,t) - g_\alpha(j,0)\big) = -\sum_{j \in \Lambda^{\mathrm{c}}_N}(\nabla_N)^{\mathrm{T}}f(j) \int_0^t ds\, \mathcal{J}_\alpha(j,s) 
\end{equation}
with $^\mathrm{T}$ denoting the transpose, i.e., $(\nabla_N)^{\mathrm{T}}f(j) = f(j-1) - f(j)$. We multiply on both sides with the same identity but now for component $\alpha'$ and test function $f(j) = \delta_{0j}$. Taking the equilibrium average and the limit $N \to \infty$ results in
\begin{multline}\label{2.10c}
\sum_{j \in \mathbb{Z}}f(j)\big(S_{\alpha\alpha'}(j,t) + S_{\alpha'\alpha}(-j,t) - 2\,S_{\alpha'\alpha}(j,0)\big) \\
= \sum_{j \in \mathbb{Z}} \Delta f(j) \int_0^t ds \int_0^t ds'\, \big\langle \mathcal{J}_\alpha (j,s); \mathcal{J}_{\alpha'}(0,s') \big\rangle_{P,\beta} \,,
\end{multline}
where $\Delta f(j) = -\nabla\nabla^{\mathrm{T}} f(j) = f(j+1) - 2f(j) + f(j-1)$ is the discrete Laplacian.
At first the identity holds only for $f$'s with integrable decay. Since, for fixed $t$, the current correlations and $S$
decay exponentially fast, one can extend to $f$'s with polynomial increase. 

For $f(j) = 1$ one arrives at
\begin{equation}\label{2.10d}
\sum_{j \in \mathbb{Z}}S_{\alpha\alpha'}(j,t) = \sum_{j \in \mathbb{Z}} S_{\alpha\alpha'}(j,0)\,. 
\end{equation}
The second moment sum rule is obtained for $f(j) = j^2$. Then
\begin{equation}\label{2.10e}
\sum_{j \in \mathbb{Z}} j^2\, \tfrac{1}{2}\big(S_{\alpha\alpha'}(j,t) + S_{\alpha'\alpha}(j,t)\big) = \int_0^t ds \int_0^t ds'\, \Gamma_{\alpha\alpha'}(s - s')\,, 
\end{equation}
using that $S_{\alpha\alpha'}(j,0) = 0$ for $j \neq 0$. With the choice $f(j) = \lvert j-i \rvert$ one has $\Delta \lvert j-i \rvert = 2\,\delta_{ij}$ and the current-current correlation can be obtained through
\begin{equation}\label{2.10f}
\sum_{j \in \mathbb{Z}} \lvert j-i\rvert \tfrac{1}{2} \big(S_{\alpha\alpha'}(j,t) + S_{\alpha'\alpha}(j,t) - 2\,S_{\alpha'\alpha}(j,0)\big) = \int_0^t ds \int_0^t ds'\, \big\langle \mathcal{J}_\alpha(i,s); \mathcal{J}_{\alpha'}(0,s') \big\rangle_{P,\beta}\,. 
\end{equation}

As discussed in \cite{Spo},
there is also a first moment sum rule, for which Eq.~\eqref{2.10b} is multiplied on both sides by $g_{\alpha'}$. Then
\begin{equation}\label{2.10g}
\sum_{j \in \mathbb{Z}} j\,S(j,t) = A C\,t\,,
\end{equation}
where $C= S(0,0)$ and $A$ is the $3\times 3$ matrix of linearized currents.

For the ring $\Lambda^\mathrm{c}_N$ the identity \eqref{2.10c} holds provided the sum is over $\Lambda^{\mathrm{c}}_N$ and $\Delta$ is replaced by $\Delta_N$. However the choice $f(j) = j^2$ results in $\Delta_N\,j^2 = 2$ for $\lvert j\rvert < N$ and $\Delta_N\,j^2 = -2 N + 1$ for $\lvert j\rvert = N$. To obtain the total current in approximation one has to choose $f(j) = j^2$ at points where the correlation is significant and use a maximally smooth interpolation otherwise. However Eq.~\eqref{2.10f} with $i=0$ has a useful finite volume version, which reads
\begin{multline}\label{2.10ga}
\sum_{j \in \Lambda^{\mathrm{c}}_N}\Big(\lvert j \rvert - \frac{1}{2N} j^2\Big) \tfrac{1}{2}\big(S_{\alpha\alpha'}(j,t) + S_{\alpha'\alpha}(j,t)\big) + \frac{1}{2N} \int_0^t ds \int_0^t ds'\, \Gamma_{\alpha\alpha'}(s-s') \\
= \int_0^t ds \int_0^t ds'\, \big\langle \mathcal{J}_\alpha(0,s); \mathcal{J}_{\alpha'}(0,s') \big\rangle_{P,\beta}\,.
\end{multline}

\noindent\textit{(ii) limit $t \to \infty$}. $\Gamma_{\alpha\alpha'}(0)$ is easily computed. But one can also determine the
long-time limit, for which the left hand side of \eqref{2.10e} is dominated by the sound peaks at $\pm ct$. The asymptotic contribution is $(ct)^2$ times the area under the peak. This is the Landau-Placzek ratio.
Using the notation from \cite{Spo} and $R^{-\mathrm{T}} = (R^{-1})^{\mathrm{T}}$, the ratio is time-independent and given by
\begin{equation}\label{2.10h}
R^{-1}\begin{pmatrix}
1 & 0 & 0\\
0 & 0 & 0\\
0 & 0 & 1
\end{pmatrix}
R^{-\mathrm{T}} = \frac{1}{m c^2\beta}
\begin{pmatrix}
1  & 0 & -P\\
0  & (m c)^2 &  0\\
-P & 0 & P^2
\end{pmatrix}\,.
\end{equation}
 By antisymmetry in $t$ the off-diagonal terms in the second moment sum rule are identical zero. Hence only the long-time limit of the diagonal matrix elements can be obtained as
\begin{equation}\label{2.21}
\lim_{t \to \infty} \Gamma_{22}(t) = \frac{m c^2}{\beta}\,,\quad \lim_{t \to \infty} \Gamma_{33}(t) = \frac{P^2}{m \beta}\,.
\end{equation}
Of course, the constant $11$ matrix element is also properly reproduced.

As an independent check, and also to fill in the off-diagonal terms, we apply the \emph{hydrodynamic projection method}, which is based on postulating that the zero modes in the space of fluctuation fields
are given by the conserved fields. We work directly in infinite volume. Fluctuation observables are of the form
\begin{equation}\label{2.22}
\xi_N (F) = \frac{1}{\sqrt{2N+1}} \sum_{j=-N}^NF_j\,.
\end{equation}
Here $F$ is a local observable on phase space, in the sense that it depends only on $r_i, p_i$ with $\lvert i\rvert \leq \mathsf{r}$, $\mathsf{r}$ the range of $F$. We denote by $F_j$ the function $F$ shifted by $j$, i.e., $F_j\big(\{r_i,p_i\}_
{\lvert i \rvert \leq \mathsf{r}+\lvert j \rvert}\big) = F\big(\{r_{i+j}, p_{i+j}\}_{ \lvert i\rvert \leq \mathsf{r}}\big)$. Thus $F_0 = F$. Also $\langle F \rangle_{P,\beta} =0$ is assumed throughout.
To give an example, for $F = F_0 = r_2\,(p_5)^2 - \langle r_0\,(p_2)^2 \rangle_{P,\beta}$ the function shifted by $j$ is $F_j = r_{2+j}\,(p_{5+j})^2 - \langle r_0\,(p_2)^2 
\rangle_{P,\beta}$.
 The asymptotic covariance,
\begin{equation}\label{2.23}
\lim_{N \to \infty} \big\langle \xi_N(F)\, \xi_N(G) \big\rangle_{P,\beta} = \sum_{j=-\infty}^\infty \langle F_j G_0\rangle
= \llangle F,G \rrangle\,,
\end{equation}
defines a Hilbert space with scalar product $\llangle \cdot,\cdot\rrangle$. Note that any difference function, $F_i - F_0$, is a null vector in this Hilbert space. If one starts from a local observable $F$, then for any $t>0$, it will depend on all $\{q_j,p_j\}_{j \in \mathbb{Z}}$. However the dependence on far away arguments is exponentially small in the distance from the origin. Such functions are called quasilocal. The sum in \eqref{2.23} has an exponential bound and the definition of
$\llangle \cdot,\cdot\rrangle$ extends to such quasilocal functions. Next note that
$\llangle F,\vec{g}(t)\rrangle = \llangle F,\vec{g}\rrangle$, since $\vec{g}(t)$ is conserved. The assumption that there are no further conservation laws translate more formally into the property 
that the time-invariant subspace is already exhausted by the local functions $\vec{g}$. 
Thus, for the long-time limit, one has to compute the projection of the currents onto the conserved fields in our Hilbert space.

For this purpose we define
\begin{equation}\label{2.24}
B_{\alpha\beta} = \llangle g_\alpha - \langle g_\alpha \rangle, \mathcal{J}_\beta - \langle \mathcal{J}_\beta \rangle\rrangle\,,\quad
C_{\alpha\beta} = \llangle g_\alpha - \langle g_\alpha \rangle, g_\beta - \langle g_\beta \rangle\rrangle\,.
\end{equation}
The hydrodynamic projection of the currents onto the conserved fields is then $B^\mathrm{T} C^{-1} B$ and thus
\begin{equation}\label{2.25}
\lim_{t \to \infty} \Gamma(t) = B^\mathrm{T}C^{-1}B
\end{equation}
as $3\times 3$ matrix. $C^{-1}$ is the normalization of the projection onto the conserved fields. For $B$ one obtains after a few partial integrations
\begin{equation}\label{2.26}
B = -\begin{pmatrix}
0 & \langle r_0;V'(r_0)\rangle& 0\\
\tfrac{1}{m} \langle p_0;p_0\rangle & 0 & \tfrac{1}{m} \langle p_0;p_0 V'(r_0)\rangle\\
0& \langle e_0;V'(r_0) \rangle & 0
\end{pmatrix}
= \frac{1}{\beta}
\begin{pmatrix}
0 & -1 & 0\\
-1 & 0 & P\\
0 &P & 0
\end{pmatrix}\,.
\end{equation}
$C$ and its inverse are is easily computed. In working out the right hand side of \eqref{2.25}, one uses that
$c^2 = \frac{\beta}{m} \langle e_0 + P r_0; e_0 + P r_0\rangle/(\langle r_0;r_0\rangle\langle e_0;e_0\rangle - \langle r_0; e_0\rangle^2)$. Then
\begin{equation}\label{2.27}
B^\mathrm{T} C^{-1} B
= \frac{1}{m \beta}
\begin{pmatrix}
1 & 0 & -P\\
0 & (m c)^2 & 0\\
-P &0 &P^2
\end{pmatrix}
\end{equation}
Thus we obtain agreement with the result from \eqref{2.21} and also identified the off-diagonal limits. For later convenience we define
\begin{equation}
\Gamma^{\scriptscriptstyle\Delta}(t) = \Gamma(t) - B^\mathrm{T} C^{-1} B\,,
\end{equation}
such that $\lim_{t \to \infty} \Gamma^{\scriptscriptstyle\Delta}(t) = 0$.

It is common practice to take care of the constant term by a redefinition of the microscopic total currents
\cite{Gr1960}. This can be done also here with the result
\begin{equation}\label{2.6b}
\vec{\mathcal{J}}^\mathrm{r}_{\mathrm{tot},\Lambda_N}(t) = 
\vec{\mathcal{J}}_{\mathrm{tot},\Lambda_N}(t) - \frac{1}{\sqrt{N}} \sum_{j=0}^{N-1} B^{\mathrm{T}} C^{-1} \vec{g}(j)
\end{equation} 
The variance of $\vec{\mathcal{J}}^\mathrm{r}(t)$ tends to $0$ as $t \to \infty$.\medskip

\noindent\textit{(iii) equivalence of ensembles}. In the microcanonical ensemble the total conserved fields have a fixed value proportional to $2N+1$, which results in the probability distribution
\begin{equation}\label{2.5}
\frac{1}{Z_N^\mathrm{mic}} \delta\bigg(\sum_{j =-N}^N r_j - \ell (2N+1)\bigg) \delta\bigg(\sum_{j =-N}^N p_j \bigg)
\delta\bigg(\sum_{j=-N}^N e_j - \mathsf{e} (2N+ 1)\bigg) \prod_{j=-N}^N dr_j\,dp_j\,.
\end{equation} 
In principle one should also constrain the momentum to $\mathsf{u} N$, which however can be shifted to $\mathsf{u} = 0$
by a Galilean transformation. Microcanonial averages carry the superscript ``mic''.
If $F_0$ is a local function, the equivalence of ensembles tells us that
\begin{equation}\label{2.7}
\lim_{N\to \infty}\langle F_0 \rangle^\mathrm{mic}_{\ell,e,\Lambda^\mathrm{c}_N} = \langle F_0 \rangle_{P,\beta} \,,
\end{equation} 
provided the parameters $(\ell,\mathsf{e})$ and $(P,\beta)$ are linked through the rules of thermodynamics. Such equivalence 
no longer holds for nonlocal observables, such as the fluctuation observable \eqref{2.22}. But the required modifications
follow a simple rule. Let us start from the finite volume correlation function $\langle F_j F_i\rangle ^{\mathrm{mic}}_{\Lambda^\mathrm{c}_N}$ with subtraction of averages understood. Then the microcanonical constraint is reflected by a uniform shift of order $1/N$ as
\begin{equation}\label{2.7a}
\langle F_jF_i\rangle ^{\mathrm{mic}}_{\Lambda^\mathrm{c}_N} \simeq \langle F_jF_i\rangle _{\Lambda^\mathrm{c}_N} + \frac{c_0}{2N+1}
\end{equation} 
with a yet to be determined constant $c_0$.
Hence
\begin{equation}\label{2.7b}
\lim_{N\to \infty}\big\langle \xi_N(F_0)^2\big\rangle^{\mathrm{mic}}_{\Lambda^\mathrm{c}_N} = \sum_{j \in \mathbb{Z}}\langle F_jF_0\rangle 
+c_0\,.
\end{equation} 
The coefficient $c_0$ can be obtained from relations between microcanonical and canonical thermodynamic potentials. In our case the microcanonical ensemble has three constraints and
\begin{equation}\label{2.7c}
\lim_{N\to \infty}\big\langle \xi_N(F_0)\, \xi_N(G_0)\big\rangle ^{\mathrm{mic}}_{\Lambda^\mathrm{c}_N}\ = 
\llangle F_0,G_0\rrangle - \sum_{\alpha,\alpha'=1}^3 \llangle F_0,g_\alpha(0)-\langle g_\alpha\rangle\rrangle (C^{-1})_{\alpha\alpha'} \llangle g_{\alpha'}(0)-\langle g_{\alpha'}\rangle,G_0\rrangle\,.
\end{equation}

Taking the microcanonical average in \eqref{2.11} defines the total current correlation $ $ $\Gamma_{N,\alpha\alpha'}^{\mathrm{mic}}(t)$. For fixed $t$ the currents are quasilocal functions and \eqref{2.7c} can still be used. Since $g_\alpha$ is conserved, the correction term is independent of $t$ and
\begin{equation}\label{2.7d}
\lim_{N \to \infty} \Gamma_{N} ^{\mathrm{mic}}(t) = \Gamma^{\scriptscriptstyle\Delta}(t)
\end{equation} 
for all $t$. If the finite system is mixing on the hypersurface defined by constant $\ell,\mathsf{u},\mathsf{e}$, then 
\begin{equation}\label{2.7e}
\lim_{t\to \infty} \Gamma_{N} ^{\mathrm{mic}}(t) = 0\,.
\end{equation} 
In principle there could be a delicate interchange between large $N$ and $t$. But Eq.~\eqref{2.7d} tells us that the decay to zero is preserved as $N \to \infty$.\medskip

\noindent\textit{Remark}. The long-time asymptotics of $\Gamma(t)$ depends on the ensemble used. Early 2000 there have been some discussions on the physical relevance of such asymptotics \cite{CamPro2000, Pro2005}. We merely add that, in the context of classical fluids, it is well understood that transport coefficients are related to the broadening of the peaks \cite{RedeL}. One dimension makes no exception, only that the broadening can be superdiffusive. The Green-Kubo formula attempts to capture the peak broadening. $\Gamma(\infty) \neq 0$ signals that there is a remnant of the ballistic motion of the sound peaks. Hence $\Gamma(\infty)$ has to be subtracted as comes out naturally in Eq.~\eqref{2.7d}. The more microscopic option is to modify the currents as explained in \eqref{2.6b}.

\section{Long-time limit}\label{sec3}

After dealing with generalities, we arrive at the real task, namely the study of the decay of the correlations
$\Gamma^{\scriptscriptstyle\Delta}_{22}(t)$, $\Gamma^{\scriptscriptstyle\Delta}_{23}(t)$ and $\Gamma^{\scriptscriptstyle\Delta}_{33}(t)$. Currently only two methods seem to be available. The first one uses the second moment sum rule together
with a highly educated, and numerically well confirmed, guess on the structure of $S(j,t)$, where we refer to
\cite{Spo,MS14}. The alternative route is to use the mode-coupling approximation from \cite{vB,Spo}, for which the current correlation equals the memory kernel. The off-diagonal matrix element can be handled only through mode-coupling, since 
by antisymmetry in $j$, respectively $t$, both sides of the second moment sum rule vanish.\medskip

\noindent\textit{Second moment sum rule}. For the diagonal matrix elements we have
\begin{equation}\label{3.2}
\frac{d^2}{dt^2}\sum_{j \in \mathbb{Z}} j^2\, S_{\alpha\alpha} (j,t) = 2\, \Gamma_{\alpha\alpha}(t) \,.
\end{equation}
Using the notation from \cite{Spo}, after normal mode transformation one obtains the diagonal form
\begin{equation}\label{3.3}
RS(j,t)R^{\mathrm{T}} = S^\sharp (j,t)\simeq \mathrm{diag}\big(f_{-1}(j,t), f_0(j,t), f_{1}(j,t)\big)
\end{equation}
valid approximately for large $j,t$. The entries are of self-similar form
\begin{align}
\label{3.4}
f_\sigma (x,t) &= (\lambda_\mathrm{s} t)^{-2/3} f_\mathrm{KPZ}\big((\lambda_\mathrm{s} t)^{-2/3}(x - \sigma ct)\big)\,,\\
\label{3.4a}
f_0(x,t) &= \chi(\{\lvert x\rvert \leq c t\}) (\lambda_\mathrm{h} t)^{-3/5}f_{\text{L\'evy},5/3}\big((\lambda_\mathrm{h} t)^{-3/5}x\big)\,,
\end{align}
$\sigma = \pm 1$. The scaling form \eqref{3.4} is based on the assumption that the modes decouple for large $t$. Then the sound mode is governed by the one-dimensional Kardar-Parisi-Zhang equation, for which the steady covariance is known exactly \cite{FeSp05,SaIm13,BoCoFe14}. The scaling form \eqref{3.4a} results from the mode-coupling approximation which yields the symmetric L\'evy stable distribution with exponent $\alpha = 5/3$, defined through its Fourier transform $\hat{f}_{\text{L\'evy},\alpha}(k) = \exp[-\lvert k \rvert^\alpha]$. $f_{\text{L\'evy},5/3}$ has a slow power law decay. However, as well confirmed by MD simulations, the scaling function is cut off beyond the sound cone $\pm ct$, which is taken into account by the indicator 
 function, $\chi$, in front of $f_{\text{L\'evy},5/3}$. 

One has to compute the second moment and differentiate twice in $t$ with the result
\begin{equation}\label{3.5}
\frac{d^2}{dt^2} \int dx\, x^2 f_\sigma(x,t) = 2\,c^2 + \tfrac{4}{9}\lambda_\mathrm{s}^2
\int dx\, x^2 f_\mathrm{KPZ}(x) (\lambda_\mathrm{s} t)^{-2/3} = 2 c^2 + 2 b_{\mathrm{s}}(\lambda_\mathrm{s} t)^{-2/3} \,,
\end{equation}
where we used that $f_{\mathrm{KPZ}}$ is even and normalized to $1$. Using $f_{\text{L\'evy},5/3}(x) \simeq \pi^{-1} \lvert x \rvert^{-8/3}$ for large $\lvert x \rvert$, one concludes that the second moment is dominated by the tails. Hence
\begin{equation}\label{3.6}
\frac{d^2}{dt^2} \int_{-c t}^{c t} dx\, x^2 f_0(x,t) \simeq \tfrac{8}{3\pi}( \lambda_\mathrm{h})^{5/3} c^{1/3} (\lambda_\mathrm{h} t)^{-2/3} = 2 b_{\mathrm{h}} (\lambda_\mathrm{h}t)^{-2/3} \,.
\end{equation}
Next we use $R C R^\mathrm{T} = \mathbbm{1}$ and the identity \eqref{2.10h} to revert the transformation in \eqref{3.3}. The subtraction terms balance exactly and, in this approximation,
\begin{equation}\label{3.7}
\Gamma^{\scriptscriptstyle\Delta}_{11}(t) = (m c^2 \beta)^{-1} \big( b_{\mathrm{s}}(\lambda_\mathrm{s} t)^{-2/3} - b_{\mathrm{h}}(\lambda_\mathrm{h}t)^{-2/3}\big) + b_{\mathrm{h}}(\lambda_\mathrm{h}t)^{-2/3} \langle r_0;r_0\rangle\,.
\end{equation}
The true $\Gamma^{\scriptscriptstyle\Delta}_{11}(t)$ does not depend on time. Thus there is a discrepancy of order $t^{-2/3}$. For momentum and energy current correlations one obtains the asymptotics
\begin{equation}\label{3.8}
\begin{split}
\Gamma^{\scriptscriptstyle\Delta}_{22}(t) &= m \beta^{-1} b_{\mathrm{s}}(\lambda_\mathrm{s} t)^{-2/3}\,, \\
\Gamma^{\scriptscriptstyle\Delta}_{33}(t) &= (m c^2 \beta)^{-1} P^2\, b_{\mathrm{s}}(\lambda_\mathrm{s} t)^{-2/3} + \big(\langle e_0;e_0\rangle - (m c^2 \beta)^{-1} P^2\big) b_{\mathrm{h}}(\lambda_\mathrm{h}t)^{-2/3}\,.
\end{split}
\end{equation}
Both total current correlations are predicted to decay as $t^{-2/3}$.

The second moments are sensitive to small changes in the scaling function. For example in \eqref{3.5} a small asymmetry of the scaling function would induce a correction of order $t^{-1/3}\times$(first moment). Also in \eqref{3.6} the second moment of
$f_{\text{L\'evy},5/3}$ diverges which makes the power law sensitive to the cut-off. For these reasons 
we regard \eqref{3.5} -- \eqref{3.8} as a sort of consistency check for the mode-coupling theory.\medskip

\noindent\textit{Mode-coupling theory}. In \cite{Spo} HS proposed mode-coupling equations. Their memory kernel is proportional to the current-current correlations with the asymptotic value already subtracted. In the following we will freely use the notation from \cite{Spo}. Referring to Eq.~(5.6) of \cite{Spo}, one has the following identity,
\begin{equation}\label{60}
\big(R \Gamma^{\scriptscriptstyle\Delta}(t) R^{\mathrm{T}}\big)_{\alpha\alpha'} = \Gamma^{\scriptscriptstyle\Delta \sharp}_{\alpha\alpha'}(t) \simeq \int_{\mathbb{R}} dx\, M_{\alpha\alpha'}(x,t)
= 2 \int_{\mathbb{R}} dx\, \mathrm{tr}\big[S^{\sharp\phi} (x,t)^\mathrm{T} G^\alpha S^{\sharp\phi} (x,t) G^{\alpha'}\big].
\end{equation}
The first equality is a definition and the $\simeq$ invokes the mode-coupling approximation. In fact, \eqref{60} coincides  with the mode-coupling formula (14a) of Ernst, Hauge, and van Leeuwen \cite{Ernst1976} when translated to the present context. Using the definition
\begin{equation}
G^{\alpha} = \tfrac{1}{2} \sum_{\alpha'=1}^3 R_{\alpha \alpha'} R^{-\mathrm{T}} H^{\alpha'} R^{-1}
\end{equation}
for the $G$ couplings (Eq.~(3.22) of \cite{Spo}) results in
\begin{equation}
\Gamma^{\scriptscriptstyle\Delta}_{\alpha\alpha'}(t) \simeq \tfrac{1}{2} \int_{\mathbb{R}} dx\, \mathrm{tr}\big[ S^{\sharp\phi}(x,t)^\mathrm{T} R^{-\mathrm{T}} H^{\alpha} R^{-1} S^{\sharp\phi}(x,t) R^{-\mathrm{T}} H^{\alpha'} R^{-1} \big].
\end{equation}
The entries of the matrices $(R^{-\mathrm{T}} H^{\alpha} R^{-1})_{\sigma\sigma'} = \langle\psi_{\sigma}\vert H^{\alpha}\vert\psi_{\sigma'}\rangle$ are provided in the Appendix A.3 of \cite{Spo}. Employing the diagonal approximation \eqref{3.3} and that, for $\alpha \neq \alpha'$, the product $f_\alpha(x,t) f_{\alpha'}(x,t)$ has vanishing overlap for large $t$, one arrives at
\begin{equation}
\Gamma^{\scriptscriptstyle\Delta}_{\alpha\alpha'}(t) \simeq \int_{\mathbb{R}} dx\, \mathrm{diag}\!\left(0, \tfrac{1}{2} \langle\psi_0, H^{\mathsf{u}} \psi_0\rangle^2 \, f_0(x,t)^2 + \langle\psi_1, H^{\mathsf{u}} \psi_1\rangle^2 \, f_1(x,t)^2, \big(\beta^{-1} c\, f_1(x,t)\big)^2 \right).
\end{equation}
Note that now only the square of the scaling function appears in the integrals, which is more robust than the second moment. The integrals are easily obtained,
\begin{equation}\label{3.12}
\int dx\, f_1(x,t)^2 = (\lambda_{\mathrm{s}}t)^{-2/3} \int dx\, f_{\mathrm{KPZ}}(x)^2\,,\quad 
\int dx\, f_0(x,t)^2 = (\lambda_{\mathrm{h}}t)^{-3/5} \int dx\, f_{\text{L\'evy},5/3}(x)^2\,. 
\end{equation}
In principle one should retain the cutoff at $\pm c t$ for $f_0$, but now this contribution is subdominant. Therefore
\begin{multline}\label{3.14}
\Gamma^{\scriptscriptstyle\Delta}(t) = \bigg( \tfrac{1}{2}(\lambda_{\mathrm{h}}t)^{-3/5} 
\langle \psi_0,H^{\mathsf{u}} \psi_0\rangle^2 \int dx\, f_{\text{L\'evy},5/3}(x)^2
 + (\lambda_{\mathrm{s}}t)^{-2/3} \langle \psi_1,H^{\mathsf{u}} \psi_1\rangle^2 \int dx\, f_{\mathrm{KPZ}}(x)^2 \bigg) \\
\times
\begin{pmatrix}
0&0&0\\
0&1&0\\
0&0&0
\end{pmatrix}\,+\,\frac{c^2}{\beta^2}(\lambda_{\mathrm{s}}t)^{-2/3} \int dx\, f_{\mathrm{KPZ}}(x)^2
\begin{pmatrix}
0&0&0\\
0&0&0\\
0&0&1
\end{pmatrix}\,.
\end{multline}
As an interesting qualitative prediction, $\Gamma^{\scriptscriptstyle\Delta}_{23}(t) = 0$ which should be interpreted as being much smaller than the diagonal terms. In fact, the MD simulations shown below confirm such prediction. $\Gamma^{\scriptscriptstyle\Delta}_{33}(t) \simeq t^{-2/3}$ in accordance with the sum rule argument. On the other hand for $\Gamma^{\scriptscriptstyle\Delta}_{22}(t)$ one obtains the two distinct power laws, $\frac{2}{3}$ and $\frac{3}{5}$. Our MD simulations seem to favor the results from mode-coupling. Note that if $\langle \psi_0,H^{\mathsf{u}} \psi_0\rangle = 0$, then $\Gamma^{\scriptscriptstyle\Delta}_{22}(t) \sim t^{-2/3}$ to leading order.

From the figures displayed below, on might conjecture that the current correlations are decreasing in time. A simple consistency check is to consider $\Gamma_{\alpha\alpha}(0) - \Gamma_{\alpha\alpha}(\infty)$. For $\alpha = 2$ this quantity seems to have either sign and monotonicity could be violated, since according to mode-coupling the asymptotic value is approached from above. On the other hand 
\begin{equation}\label{3.15}
\Gamma_{33}(0) - \Gamma_{33}(\infty) = (m \beta)^{-1}\big(\langle V'^2 \rangle_{P,\beta}
- \langle V' \rangle_{P,\beta}^2\big) \geq 0\,,
\end{equation}
which supports monotone decrease.\medskip

\noindent\textit{A comment on early literature}. Pomeau and R\'esibois \cite{PoRe75}, p.~118, predict the $t^{-2/3}$ decay of the total current correlation for a one-dimensional system with a single conserved field through a self-consistency relation for the exponents. In our language this corresponds to the stochastic Burgers equation which was studied then in greater detail by Forster, Nelson, and Stephen \cite{FoNeSt77}. On the MD side, in 1998 Lepri, Livi, and Politi \cite{LeLiPo98} report on the total heat current correlation for the FPU $\beta$ chain and Hatano \cite{Ha98} for the Toda chain with alternating masses. For a hard point system with alternating masses the $t^{-2/3} $ decay is well documented in \cite{GrNa02}.

\section{Molecular dynamics current correlations for hard collision models}\label{sec4}

In Ref.~\cite{MS14} we studied numerically the correlator $S(j,t)$ for three models with hard collisions, that is, with a piecewise constant interaction potential and such that the order of particles is maintained, $r_j \geq 0$ for all $j$. We continue this program with the total current correlations, which require separate MD simulations. For details of the method we refer to \cite{MS14}. The number of lattice sites is chosen to be $N = 4096$ throughout. The maximal time is $t = 1024$ and the sound speed ranges from $1.732$ to $1.744$. We recall the models from~\cite{MS14} and discuss for each one separately the results for the current correlations.\medskip

\begin{figure}[!ht]
\centering
\includegraphics[width=0.4\textwidth]{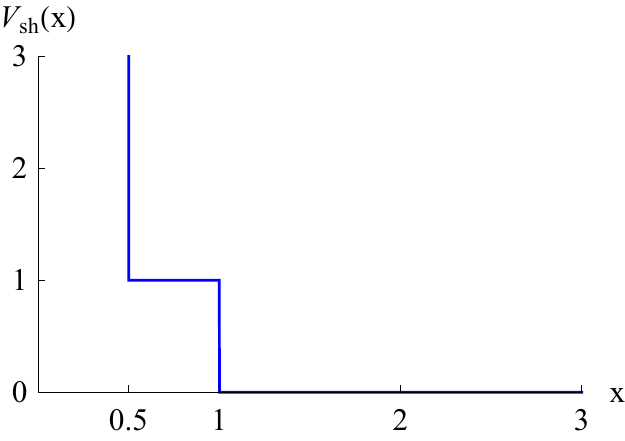}
\caption{(Color online) Shoulder interaction potential defined in Eq.~\eqref{eq:Vshoulder}.}
\label{fig:Vshoulder}
\end{figure}

\noindent\textit{(i) Monoatomic chain with ``shoulder'' interaction potential.} The potential is defined by
\begin{equation}\label{eq:Vshoulder}
V_\mathrm{sh}(x) = \begin{cases}
\infty & \mathrm{for}\quad \lvert x \rvert \leq \tfrac{1}{2} \,, \\
1      & \mathrm{for}\quad \tfrac{1}{2} < \lvert x \rvert < 1 \,,\\
0      & \mathrm{for}\quad 1 \leq \vert x \rvert \,,
\end{cases}
\end{equation}
as visualized in Fig.~\ref{fig:Vshoulder}.

The momentum and energy currents are a sequence of $\delta$-spikes at each time of collision. 
To compute the weight of the $\delta$-spike we assume that the particles $j$ and $j+1$ collide during the time interval $[t,t']$, which is taken short enough such that there are no collisions with neighboring particles. Then for the momentum current spike one obtains
\begin{equation}\label{J2collision}
\begin{split}
\int_t^{t'} ds\,\mathcal{J}_2(j+1,s) &= - \int_t^{t'} ds\,V'(r_j(s)) \\
&= \int_t^{t'} ds\,\frac{d}{ds} p_{j+1}(s) 
-\int_t^{t'} ds\, V'(r_{j+1}(s)) = p_{j+1}' - p_{j+1}\,,
\end{split}
\end{equation}
since by assumption the potential at $r_{j+1}$ is constant within the interval $[t,t']$. Here $p_{j+1}'$ is the momentum \emph{after} the collision. By momentum conservation one can also switch to the momentum transfer for particle $j$ through $p_{j+1}' - p_{j+1}= -p_{j}' + p_{j}$. Likewise the weight of a $\delta$-spike for the energy current is obtained as
\begin{equation}\label{J3collision}
\begin{split}
&\int_t^{t'} ds\,\mathcal{J}_3(j+1,s) = - \int_t^{t'} ds\,\tfrac{1}{m} p_{j+1}(s)\, V'(r_j(s)) \\
&= \int_t^{t'} ds\,\tfrac{1}{m} p_{j+1}(s) \Big( \frac{d}{ds} p_{j+1}(s) - V'(r_{j+1}(s)) \Big) = \tfrac{1}{2 m} p_{j+1}'^2 - \tfrac{1}{2 m} p_{j+1}^2 \,.
\end{split}
\end{equation}
One has to distinguish between the hard collision at $r_j = \frac{1}{2}$ and the collision at the potential step at $r_j = 1$. We refer to \cite{MS14} for the concrete formulas of $p_j'$ and $p_{j+1}'$.

\begin{figure}[!ht]
\centering
\subfloat[momentum current correlations]{
\includegraphics[width=0.31\textwidth]{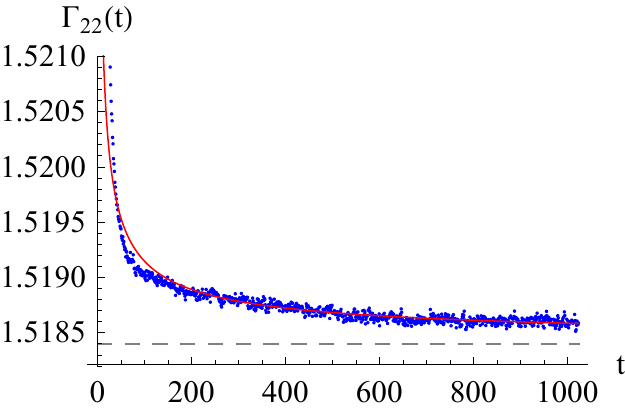}}
\hspace{0.01\textwidth}
\subfloat[logarithmic plot of $\Gamma^{\scriptscriptstyle\Delta}_{22}(t)$]{
\includegraphics[width=0.31\textwidth]{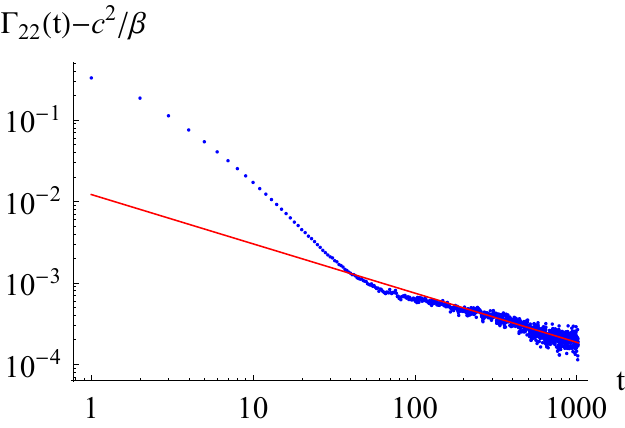}}
\hspace{0.01\textwidth}
\subfloat[double time-integrated $\Gamma^{\scriptscriptstyle\Delta}_{22}(t)$]{
\includegraphics[width=0.31\textwidth]{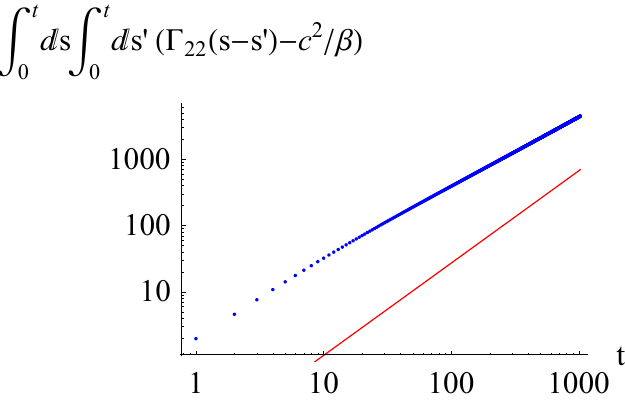}}\\
\subfloat[energy current correlations]{
\includegraphics[width=0.31\textwidth]{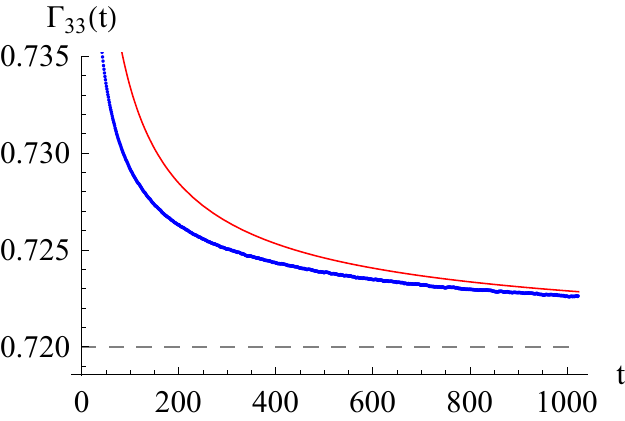}}
\hspace{0.01\textwidth}
\subfloat[logarithmic plot of $\Gamma^{\scriptscriptstyle\Delta}_{33}(t)$]{
\includegraphics[width=0.31\textwidth]{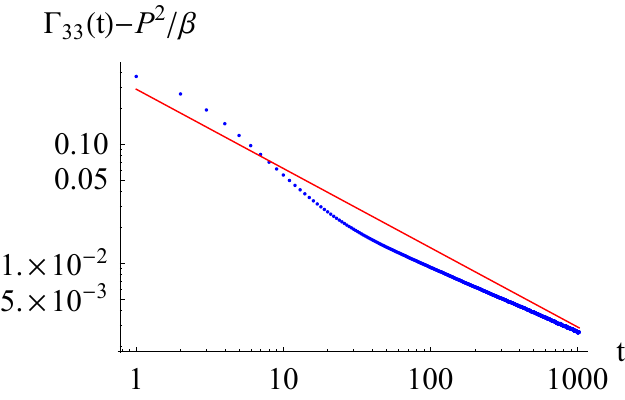}}
\hspace{0.01\textwidth}
\subfloat[double time-integrated $\Gamma^{\scriptscriptstyle\Delta}_{33}(t)$]{
\includegraphics[width=0.31\textwidth]{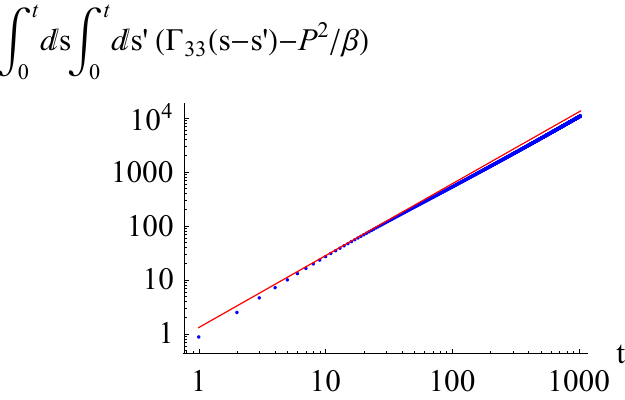}}
\caption{(Color online) Time correlations of the total momentum and energy currents (blue dots) for the monoatomic chain with shoulder interaction potential. Dashed horizontal lines show the theoretical asymptotic value, and red curves the prediction \eqref{3.14} based on mode-coupling theory.}
\label{fig:shoulder_J_corr}
\end{figure}
For each simulation run, the system is initially equilibrated according to the canonical probability density \eqref{2.6} with parameters $P = 1.2$ and $\beta = 2$. The corresponding numerical values for the sound speed, $R$ matrix, and $G$ couplings can be found in the appendix of Ref.~\cite{MS14}. Fig.~\ref{fig:shoulder_J_corr} visualizes the total momentum and energy current time correlations, including logarithmic and time-integrated plots, after averaging over $10^7$ simulation runs. $\Gamma_{22}(t)$ has been shifted by $-0.00015$ to correct for deviations from the asymptotic value due to the finite number of samples. At our parameters, $\lambda_{\mathrm{s}} = 1.0365$, $\lambda_{\mathrm{h}} = 1.7111$, $\langle\psi_0, H^{\mathsf{u}} \psi_0\rangle = 0.4029$ and $\langle\psi_1, H^{\mathsf{u}} \psi_1\rangle = 0.0533$. Inserting in the mode-coupling predictions \eqref{3.14} (red curves) one arrives at a fairly good agreement. The noise level of the momentum current correlations appears to be large compared to the energy current correlations. This is due to the smaller distance from the asymptotic value and also the different scales of the $y$-axes. The logarithmic energy correlation plot seems to suggest a power law different from $\sim t^{-2/3}$, but from other MD simulations \cite{vBPo} it is known that the energy current converges slowly. As a control check (not shown) we also measured the stretch current correlation,
which indeed merely fluctuates around $1/(m\beta)$.

\begin{figure}[!ht]
\centering
\includegraphics[width=0.4\textwidth]{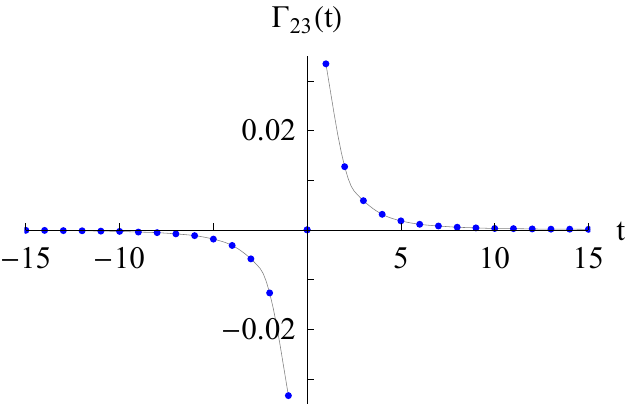}
\caption{(Color online) Time cross-correlation of the total momentum and energy currents for the monoatomic chain with shoulder interaction potential, as in Fig.~\ref{fig:shoulder_J_corr}. Note the shorter scale of the time axis.}
\label{fig:shoulderGamma23}
\end{figure}
Besides $\Gamma_{22}(t)$ and $\Gamma_{33}(t)$ we also measured the momentum-energy cross correlation $\Gamma_{23}(t)$, which is an antisymmetric function in $t$, in good agreement with the numerical data. From mode-coupling one obtains $\Gamma_{23}(t) = 0$ to leading order, implying that one should study the subleading corrections, which has not been accomplished yet. Numerically we find a very fast decay to zero (order of $10$ time units) for $\Gamma_{23}(t)$, see Fig.~\ref{fig:shoulderGamma23}.\medskip

\noindent\textit{(ii) Biatomic chain of hard-point particles}. Neighboring particles interact through elastic collisions at distance zero.
To obtain a non-integrable model the masses are alternating, where we set $m_j = 1$ for even $j$ and $m_j = 3$ for odd $j$. Using local momentum and energy conservation during a collision, the formulas \eqref{J2collision} and \eqref{J3collision} turn into
\begin{align}
\int_t^{t'} ds\,\mathcal{J}_2(j+1,s) &= -\tfrac{2 m_j m_{j+1}}{m_j + m_{j+1}} \Big( \tfrac{1}{m_{j+1}} p_{j+1} - \tfrac{1}{m_j} p_j \Big)\,, \\
\int_t^{t'} ds\,\mathcal{J}_3(j+1,s) &= \tfrac{1}{m_j + m_{j+1}} (p_j + p_{j+1}) \int_t^{t'} ds\,\mathcal{J}_2(j+1,s)\,.
\end{align}
A unit cell now consists of two neighboring lattice sites. In principle one should repeat the computations of Sect.~\ref{sec2}. But at the end the only modification for the long-time asymptotics and the mode-coupling equations amounts to replacing $m$ by the average mass $\bar{m} = \tfrac{1}{2}(m_j + m_{j+1})$.

\begin{figure}[!ht]
\centering
\subfloat[momentum current correlations]{
\includegraphics[width=0.31\textwidth]{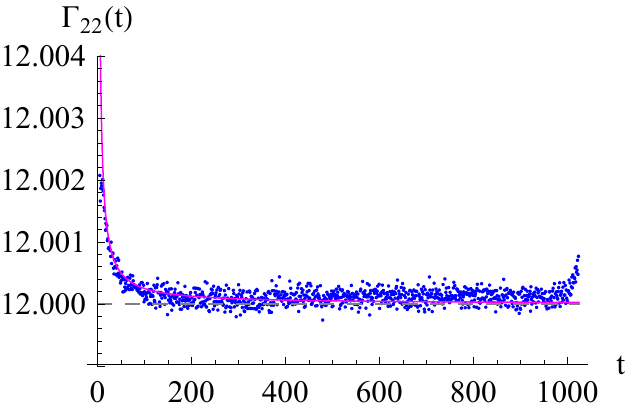}}
\hspace{0.01\textwidth}
\subfloat[logarithmic plot of $\Gamma^{\scriptscriptstyle\Delta}_{22}(t)$]{
\includegraphics[width=0.31\textwidth]{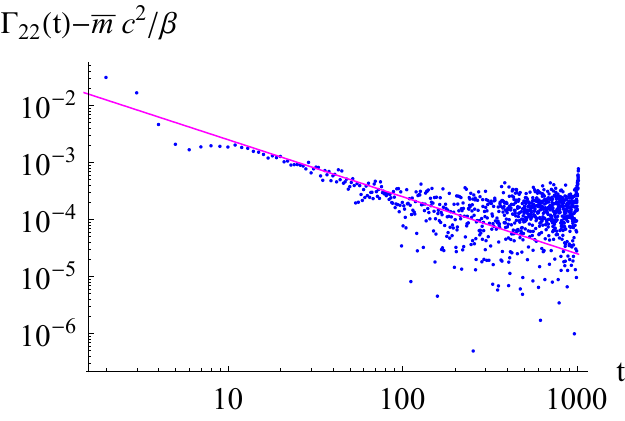}}
\hspace{0.01\textwidth}
\subfloat[double time-integrated $\Gamma^{\scriptscriptstyle\Delta}_{22}(t)$]{
\includegraphics[width=0.31\textwidth]{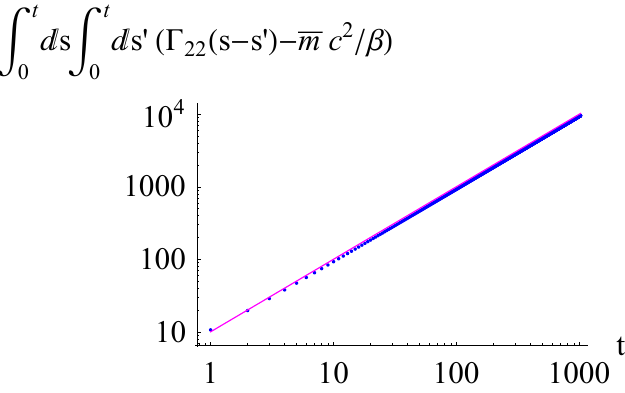}}\\
\subfloat[energy current correlations]{
\includegraphics[width=0.31\textwidth]{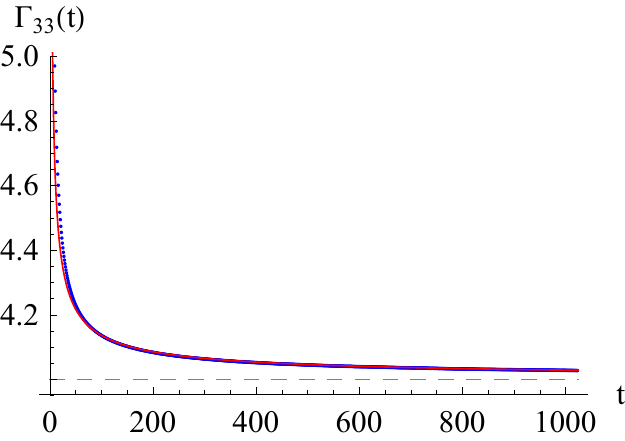}}
\hspace{0.01\textwidth}
\subfloat[logarithmic plot of $\Gamma^{\scriptscriptstyle\Delta}_{33}(t)$]{
\includegraphics[width=0.31\textwidth]{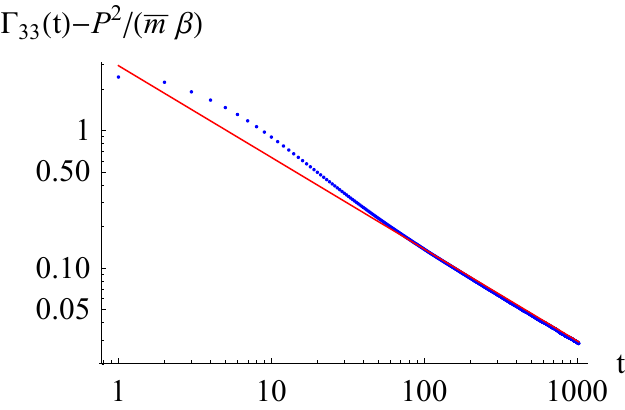}}
\hspace{0.01\textwidth}
\subfloat[double time-integrated $\Gamma^{\scriptscriptstyle\Delta}_{33}(t)$]{
\includegraphics[width=0.31\textwidth]{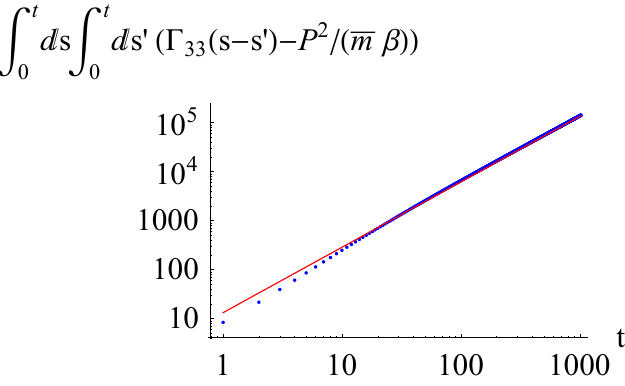}}
\caption{(Color online) Time correlations of the total momentum and energy currents (blue dots) for the biatomic chain of hard-point particles with alternating masses. The red curves in the lower row show the prediction \eqref{3.14} for the energy current correlations, and the magenta curves in (a) and (b) a heuristically fitted $\sim t^{-1}$ decay, and in (c) $\sim t$.}
\label{fig:altern_mass_J_corr}
\end{figure}
We have chosen the parameters $P = 2$ and $\beta = \frac{1}{2}$ for the canonical probability density \eqref{2.6}, as in Ref.~\cite{MS14}. Fig.~\ref{fig:altern_mass_J_corr} visualizes the total momentum and energy current time correlations after averaging over $10^7$ simulation runs. The total current is the sum over even lattice sites only, in accordance with a unit cell of two sites. $\Gamma_{22}(t)$ has been shifted by $-0.0064$ to correct for deviations due to the finite number of samples. The $t^{-2/3}$ decay of the energy current correlation predicted in \eqref{3.14} sets in considerably earlier than in the case of the shoulder potential. However, the prediction for the momentum current is not applicable since the matrix elements $\langle \psi_0, H^{\mathsf{u}} \psi_0\rangle$ and $\langle \psi_1, H^{\mathsf{u}} \psi_1\rangle$ both vanish. To arrive at a decay for the momentum current correlation, one would have to study subleading corrections. Over the time window available our data suggest a decay as $t^{-1}$, see magenta lines in the upper row of Fig.~\ref{fig:altern_mass_J_corr}. The small tip at the maximum correlation time in Fig.~\ref{fig:altern_mass_J_corr}a presumably results from colliding sound peaks due to the periodicity of the lattice.\medskip

\noindent\textit{(iii) Biatomic chain of hard-point particles with square-well potential}. The model is as described in \textit{(ii)}, but the interaction potential admits only a maximal stretch, $a$, that is
\begin{equation}
V_\mathrm{sw}(x) = 0 \quad \mathrm{for}\ 0 < \lvert x\rvert < a\,,\quad
V_\mathrm{sw}(x) = \infty\ \mathrm{otherwise}\,.
\end{equation}
In our case $a = 1$.

\begin{figure}[!ht]
\centering
\subfloat[momentum current correlations]{
\includegraphics[width=0.31\textwidth]{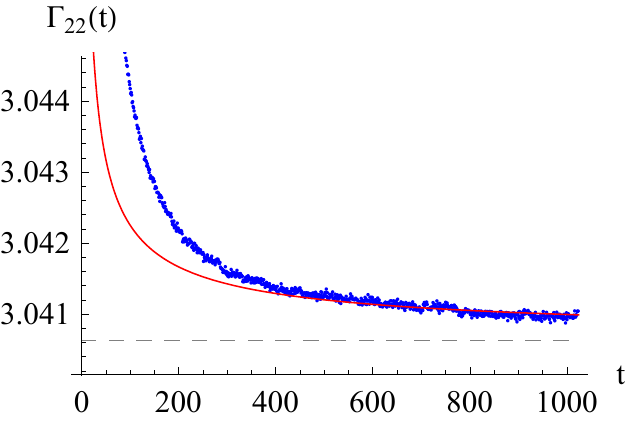}}
\hspace{0.01\textwidth}
\subfloat[logarithmic plot of $\Gamma^{\scriptscriptstyle\Delta}_{22}(t)$]{
\includegraphics[width=0.31\textwidth]{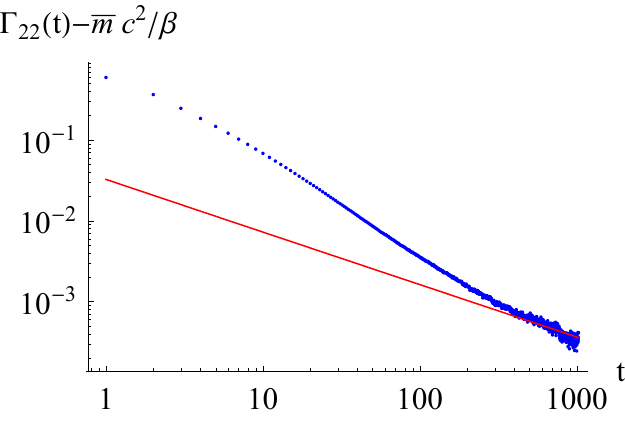}}
\hspace{0.01\textwidth}
\subfloat[double time-integrated $\Gamma^{\scriptscriptstyle\Delta}_{22}(t)$]{
\includegraphics[width=0.31\textwidth]{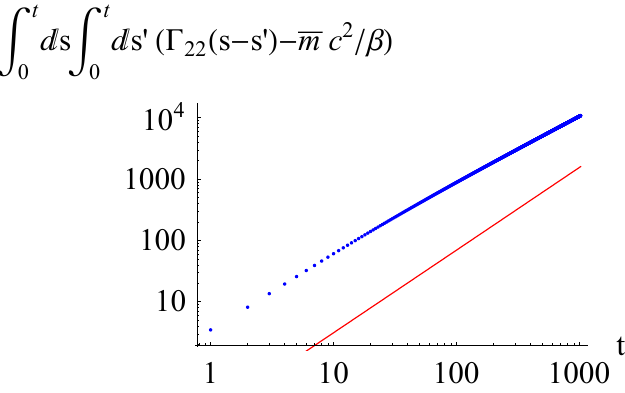}}\\
\subfloat[energy current correlations]{
\includegraphics[width=0.31\textwidth]{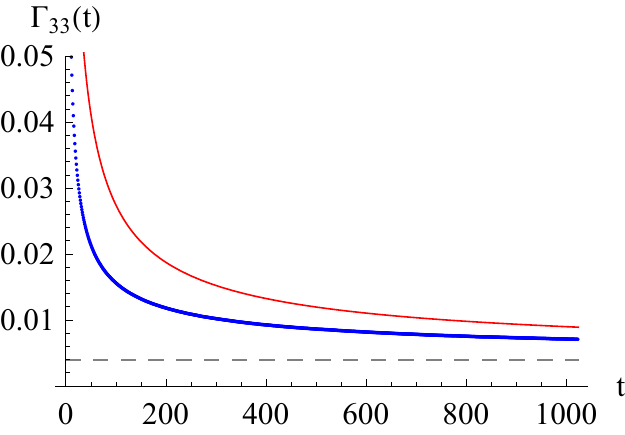}}
\hspace{0.01\textwidth}
\subfloat[logarithmic plot of $\Gamma^{\scriptscriptstyle\Delta}_{33}(t)$]{
\includegraphics[width=0.31\textwidth]{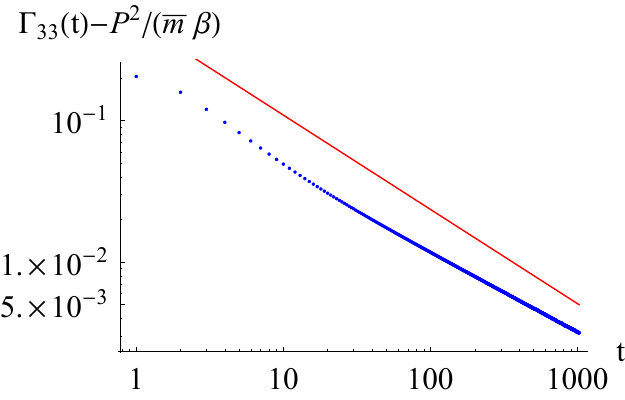}}
\hspace{0.01\textwidth}
\subfloat[double time-integrated $\Gamma^{\scriptscriptstyle\Delta}_{33}(t)$]{
\includegraphics[width=0.31\textwidth]{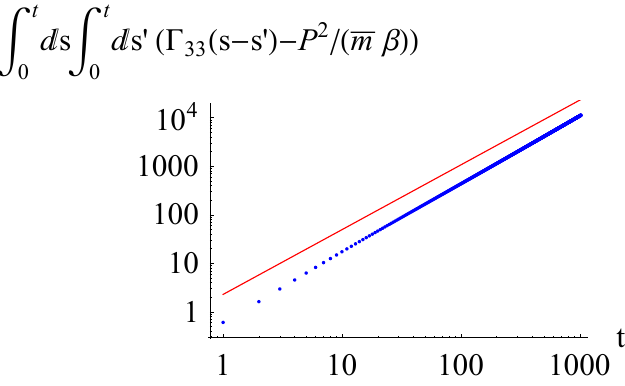}}
\caption{(Color online) Time correlations of the total momentum and energy currents (blue dots) for the biatomic chain with square-well potential. The red curves show the prediction \eqref{3.14} based on mode-coupling theory.}
\label{fig:square_well_J_corr}
\end{figure}
To have another example in the standard class, we set $P = \frac{1}{8}$ such that $G^\sigma_{\sigma\sigma}\neq 0$, and $\beta = 2$. Appendix~\ref{secG} lists the corresponding sound speed, $R$ matrix and $G$ couplings. This choice of the pressure differs from Ref.~\cite{MS14}, where we had set $P=0$ to obtain a model in the universality class with diffusive sound peaks and $\tfrac{3}{2}$-L\'evy heat peak. The momentum current correlation should then decay integrably, while the energy current correlation is predicted to decay as $t^{-1/2}$. For the present setting $P = \frac{1}{8}$, we rely on the mode-coupling prediction \eqref{3.14}. The matrix elements at our parameters have the values $\langle \psi_0, H^{\mathsf{u}} \psi_0\rangle = -0.4933$ and $\langle \psi_1, H^{\mathsf{u}} \psi_1\rangle = 0.1968$, and the nonuniversal coefficients $\lambda_{\mathrm{s}} = 0.4468$ and $\lambda_{\mathrm{h}} = 7.7382$. The total momentum and energy current time correlations $\Gamma_{22}(t)$ and $\Gamma_{33}(t)$ are shown in Fig.~\ref{fig:square_well_J_corr} after averaging over $10^7$ simulation runs and shifting $\Gamma_{22}(t)$ by $0.0006$ to correct for deviations due to the finite number of samples. Indeed $\Gamma_{22}(t)$ decays slower as compared to the model \textit{(ii)} for which the leading coefficient vanishes. While the mode-coupling prediction fits the numerical data reasonably well, one would require larger system sizes and correlation times to decisively determine the asymptotic momentum current time decay exponent from the numerical data.

\section{Current statistics and Baik-Rains distribution}\label{sec5}

The covariance of currents in equilibrium is a conventional topic of statistical mechanics.
To go beyond opens many options. One line of research is to consider the higher cumulants of time-integrated currents and their respective large deviation function \cite{Gerschenfeld, DerridaGe2010}. In this case $N$ is fixed and the time span of observation is much larger than $N/c$. Here we consider the case where first $N \to \infty$. Then a natural quantity would be the time-integrated current at a fixed site, e.g. $j = 0$,
 \begin{equation}\label{4.1}
\int_0^t ds\, \vec{\mathcal{J}}(0,s)
\end{equation}
and one is interested to study its statistics for large $t$. 
Generically one expects to have a central limit theorem with Gaussian fluctuations of order $\sqrt{t}$.
Since transport is anomalous, one might wonder whether other, judiciously chosen, currents could have more exotic statistical features. The guidance comes from nonlinear fluctuating hydrodynamics. If modes decouple, then the sound mode is described by the stochastic Burgers equation, which is the canonical case of a single-component stochastic 
field theory governed by a conservation law. So we have to first take a detour in order to explain the current fluctuations in this context. As example we will use the totally asymmetric simple exclusion process (TASEP) \cite{FeSp05}.
But the identical features have been obtained also for other models \cite{PrSp04,BoCoFe14}.

The TASEP is a model for particles on $\mathbb{Z}$, which randomly hop to the right under the constraint of at most one particle per site. A particle configuration is denoted by $\{\eta(j)\}_{j \in \mathbb{Z}}$, where $\eta(j) = 1$ if site $j$ is occupied and $\eta(j) = 0$ if site $j$ is empty. The particle configuration at time $t$ is denoted by $\eta(j,t)$. The dynamics consists of random jumps to the right. The particle at site $j$ waits an exponentially distributed time and then attempts to jump to $j+1$. If this site happens to be occupied the attempt is discarded, while if it is empty the jump is performed. All jump attempts are independent. In the steady state the $\{\eta(j)\}$'s are independent with $\mathbb{P}(\eta(j) = 1) = \rho$, $\mathbb{P}(\eta(j) = 0) = 1- \rho$, and average density $0 < \rho < 1$. As for anharmonic chains, we study the space-time stationary process $\eta(j,t)$ at fixed density $\rho$. Clearly the particle number is locally conserved. The particle current is denoted by $\tilde{\mathcal{J}}(j,t)$ and consists of a sequence of $\delta$-functions located at the times when a particle jumps from $j$ to $j+1$. In particular
\begin{equation}\label{4.2}
\int_0^t ds\, \tilde{\mathcal{J}}(j,s) = \tilde{\mathcal{J}}(j,[0,t])
\end{equation}
equals the number of jumps across the bond $(j,j+1)$ during the time span $[0,t]$. 

The analogue of \eqref{4.1} is then
 \begin{equation}\label{4.3}
\tilde{X}(0,t) = \tilde{\mathcal{J}}(0,[0,t]) - t \rho(1-\rho)\,. 
\end{equation}
$t^{-1/2}\tilde{X}(0,t)$ satisfies a central limit theorem with variance $\lvert 1 - 2\rho\rvert \rho (1 - \rho)$ \cite{FeFo94}. Only at $\rho = \tfrac{1}{2}$ the variance vanishes, which signals anomalous fluctuations. To capture other densities, one has to generalize from the current through the space-time path $s \mapsto (0,s), \,0 \leq s \leq t,$ to the $v$-path $s \mapsto (\lfloor vs\rfloor, s),\,0 \leq s \leq t$, where $\lfloor vs \rfloor$ means integer part and $v$ is a path parameter. Due to the conservation law, the two-dimensional vector field $\big(\eta(j,t),- \tilde{\mathcal{J}}(j,t)\big)$ is curl free. Hence, without modifying the current, the $v$-path can be deformed to $j \mapsto (j,0)$, $j = 1,\dots,\lfloor vt\rfloor,$ and $s \mapsto (\lfloor vt\rfloor, s)$, $0 \leq s \leq t$. Thus the current across the $v$-path is given by 
\begin{equation}\label{4.4}
\tilde{X}(\lfloor vt\rfloor,t) = \int_0^t ds\, \big( \tilde{\mathcal{J}}(\lfloor vt\rfloor,s) - \rho (1-\rho)\big) - \sum_{j=0}^{\lfloor v t\rfloor}\big(\eta(j,0) - \rho\big)\,,
\end{equation}
where steady state averages have been subtracted. To capture the anomalous fluctuations one has to choose $v$ as the velocity of propagation of a small density fluctuation relative to the background density $\rho$, in other words one has to follow the ballistic motion of the correlation peak. Hence, in our example $v = 1- 2\rho$. For PNG and TASEP, Baik and Rains \cite{BaRa00} prove that 
\begin{equation}\label{4.4a}
\lim_{t \to \infty} \mathbb{P}\big(\{t^{-1/3}\tilde{X}(\lfloor(1-2\rho)t\rfloor,t) \leq s\}\big) = F_\mathrm{BR}(s)\,. 
\end{equation}
The Baik-Rains distribution function $F_\mathrm{BR}$, denoted by $F_0$ in \cite{BaRa00} and \cite{PrSp00}, can be written in terms of the Hastings-McLeod solution of the Painlev\'e II equation.

We recall that the Hastings-McLeod solution is the unique solution of
\begin{equation}\label{4.4b}
u''(s) = 2u(s)^2 + su(s) 
\end{equation}
satisfying $u(s) < 0$ for all $s\in \mathbb{R}$. It has the asymptotics
\begin{equation}\label{4.4c}
u(s) \simeq -\mathrm{Ai} (s)\,\,\mathrm{for}\,\,s \to \infty\,,\quad
u(s) \simeq - \sqrt{-s/2}\,\,\mathrm{for}\,\,s \to -\infty\,. 
\end{equation}
The Hastings-McLeod solution is unstable and requires high precision for its numerical realization.
A tabulation, also used below, can be found on the homepage of M. Pr\"{a}hofer \cite{Pra}.
In terms of $u$ one defines the auxiliary functions
\begin{align}\label{4.4d}
U(s) &= - \int_s^\infty dx\, u(x) \,,\\
v(s) &= \big(u(s)^2 + s\big) u(s)^2 - u'(s)^2\,,\\
V(s) &=  - \int_s^\infty dx\, v(x) \,.
\end{align}
Then the Baik-Rains distribution function is defined by
\begin{equation}\label{4.4e}
F_\mathrm{BR}(s) = \left(1 - \big(s + 2u'(s) + 2u(s)^2 \big) v(s)\right) \exp\!\big[-2U(s)-V(s)\big] \,.
\end{equation}

We return to the anharmonic chain and consider the $\alpha$-current through a path from $(0,0)$ to $(y,t)$, $y \in \mathbb{Z}$. 
To simplify notation we take $y \geq 0$, but the final formulas extend to all $y$ in the obvious way. We deform the path as before and define
\begin{equation}\label{4.5}
X_\alpha(y,t) = \int_0^t ds\, \big(\mathcal{J}_\alpha(y,s) - \langle\mathcal{J}_\alpha\rangle\big) - \sum_{j=0}^{y-1} \big(g_\alpha(j,0) - \langle g_\alpha \rangle\big)\,.
\end{equation}
We are interested in the statistics of this current in the limit of large $t$ when setting $y = \lfloor vt\rfloor$.
But first we derive an identity which relates the second moment of $X_\alpha$ to the structure function $S$. 
We claim that, for infinite volume,
\begin{equation}\label{4.6}
\Upsilon_{\alpha\alpha'}(y,t) = \big\langle X_\alpha(y,t); X_{\alpha'}(y,t)\big\rangle_{P,\beta} = \sum_{j\in \mathbb{Z}}\lvert j-y\rvert\tfrac{1}{2} \big(S_{\alpha\alpha'}(j,t)
+ S_{\alpha'\alpha}(j,t)\big)\,. 
\end{equation}

To establish \eqref{4.6} we multiply out the square. This gives four contributions, the cross terms being related by exchanging 
$\alpha$ and $\alpha'$. One arrives at 
\begin{equation}\label{4.7}
\sum_{i,j=0}^{y-1}\big\langle g_\alpha(i,0); g_{\alpha'}(j,0) \big \rangle_{P,\beta} = y \sum_{j\in \mathbb{Z}} S_{\alpha\alpha'}(j,t)\,,
\end{equation}
where we used \eqref{2.10d}, and
\begin{equation}\label{4.8}
\int_0^t ds \int_0^t ds'\, \big\langle\mathcal{J}_\alpha(y,s); \mathcal{J}_{\alpha'}(y,s') \big \rangle_{P,\beta} = 
\sum_{j\in \mathbb{Z}} \lvert j \rvert \tfrac{1}{2} \big(S_{\alpha\alpha'}(j,t) + S_{\alpha'\alpha}(j,t)\big)\,,
\end{equation}
where we employed stationarity in $y$ and \eqref{2.10f} with $i = 0$. For the cross term we adopt the shorthand
$g_\alpha(f,t) = \sum_j f(j) g_{\alpha}(j,t)$ and start from the identity
\begin{equation}\label{4.9}
\big\langle g_\alpha(h,0); g_{\alpha'}(f,t) - g_{\alpha'}(f,0) \big\rangle_{P,\beta} = 
- \int_0^t ds \sum_{i,j\in \mathbb{Z}}\big\langle g_\alpha(i,0); \mathcal{J}_{\alpha'}(j,s) \big\rangle_{P,\beta}\, f(j) \nabla h(i) \,
\end{equation}
by \eqref{2.10b} for $N = \infty$ and partial integration. We set $f(j) = \delta_{jy}$ and $h(i) = 0$ for $i \leq 0$, $h(i) = i$ for $0 \leq i\leq y$, and $h(i) = y$ for $y \leq i$. Then 
\begin{equation}\label{4.10}
\sum_{i=0}^{y-1} \int_0^t ds\, \big\langle g_\alpha(i,0); \mathcal{J}_{\alpha'}(y,s)\big\rangle_{P,\beta} = \sum_{j\in \mathbb{Z}}\big({-h(y-j)} + y\big)S_{\alpha'\alpha}(j,t) \,,
\end{equation}
using \eqref{2.10d} as before. Summing \eqref{4.7}, \eqref{4.8} and subtracting the expression \eqref{4.10} and its adjoint yields \eqref{4.6}.

To focus our attention to a specific peak, we transform $\Upsilon(y,t)$ to normal modes by
\begin{equation}\label{4.11}
R\Upsilon(y,t)R^\mathrm{T} = \Upsilon^\sharp(y,t) = \sum_{j\in \mathbb{Z}} \lvert j-y \rvert \tfrac{1}{2} \big(S^\sharp(j,t) + S^{\sharp \mathrm{T}}(j,t)\big)\,.
\end{equation}
Inserting the approximation \eqref{3.3} for $S^\sharp$ one obtains
\begin{align}\label{4.12}
\Upsilon^\sharp_{11}(\lfloor ct\rfloor,t) &= \sum_{j\in \mathbb{Z}} \lvert j-\lfloor ct\rfloor\rvert\,f_{1}(j,t)
\simeq (\lambda_\mathrm{s} t)^{2/3}\int dx\, \lvert x \rvert f_{\mathrm{KPZ}}(x)\,, \\\label{4.12a}
\Upsilon^\sharp_{00}(0,t) &= \sum_{j\in \mathbb{Z}} \lvert j \rvert\, f_{0}(j,t)
 \simeq (\lambda_\mathrm{h} t)^{3/5}\int dx\, \lvert x\rvert\, f_{\text{L\'evy},5/3}(x)\,.
\end{align}
On the other hand, if $v \neq c$, then $\Upsilon^\sharp_{11}(\lfloor vt\rfloor,t)$ is proportional to $t$, as is $\Upsilon^\sharp_{00}(\lfloor vt\rfloor,t)$ whenever $v \neq 0$. Hence one would expect that in these cases
\begin{equation}\label{4.13}
t^{-1/2}\,X^\sharp_1(\lfloor vt\rfloor,t)\,,\quad t^{-1/2}\,X^\sharp_0(\lfloor vt\rfloor,t) 
\end{equation}
have a Gaussian distribution for large $t$.

If the decoupling hypothesis for the sound mode is valid in a sufficiently strong sense, then one would expect to have the same limiting distribution as for the TASEP. This means that 
\begin{equation}\label{4.14}
t^{-1/3}\,X^\sharp_1(\lfloor ct\rfloor,t)
\end{equation}
is of order 1 and has Baik-Rains as limit distribution. For the heat peak no comparable theory is available.

\section{Current statistics for hard collisions with alternating masses}\label{sec6}

For the three models discussed above we simulated the current statistics. At the level of total current correlations the alternating mass model with hard collisions has optimal convergence. The same observation holds for the current statistics and we display our results 
only for this model.

\begin{figure}[!ht]
\centering
\includegraphics[width=0.38\textwidth]{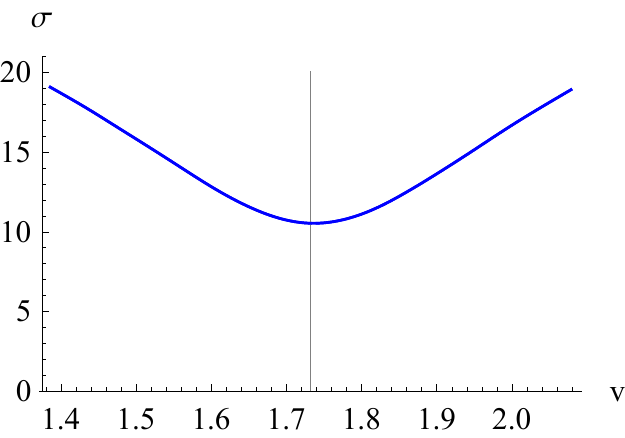}
\caption{(Color online) Standard deviation of the right sound mode current integrated along the path \eqref{4.5} with $y = \lfloor vt\rfloor$ in dependence of $v$, at $t = 1024$. The gray vertical line shows the actual sound speed $c = \sqrt{3}$.}
\label{fig:altern_mass_Jsite_vt_std_deviation}
\end{figure}
\begin{figure}[!ht]
\centering
\subfloat[left sound mode]{
\includegraphics[width=0.31\textwidth]{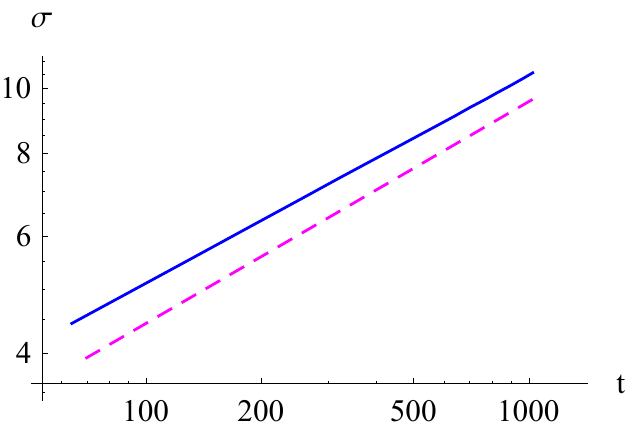}}
\hspace{0.01\textwidth}
\subfloat[heat mode]{
\includegraphics[width=0.31\textwidth]{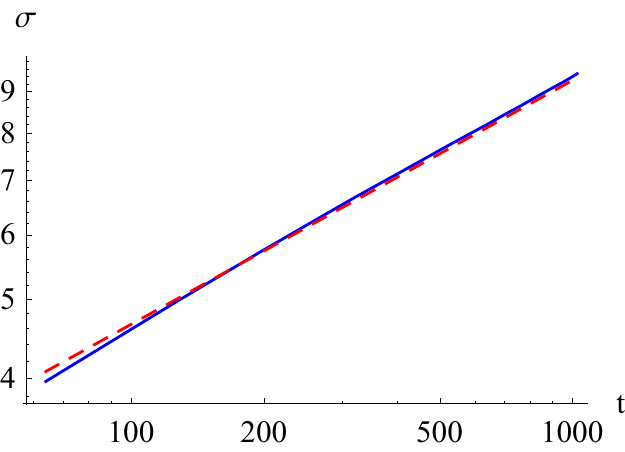}}
\hspace{0.01\textwidth}
\subfloat[right sound mode]{
\includegraphics[width=0.31\textwidth]{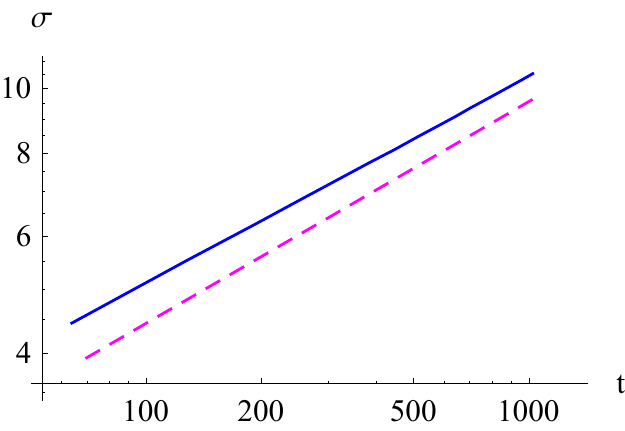}}
\caption{(Color online) Standard deviation of the normal-mode currents integrated along the path \eqref{4.5} with end points $y = -ct,0,ct$. The dashed lines show the theoretically predicted scaling $t^{1/3}$ for the sound peaks and $t^{3/10}$ for the heat peak as in Eqs.~\eqref{4.12} and \eqref{4.12a}, respectively, with $\lambda_{\mathrm{s}} = 2$ and $\lambda_{\mathrm{h}} = 0.94878$.}
\label{fig:altern_mass_Jsite_time_std_deviation}
\end{figure}

On the most basic level, from Eq.~\eqref{4.12} one concludes that if the end point for the momentum current is exactly on the appropriate sound peak, then the variance is reduced. In Fig.~\ref{fig:altern_mass_Jsite_vt_std_deviation} we display the result of the MD simulation in dependence of the end point. As expected, the standard deviation assumes its minimum at $v = c$. Without dispersion one should see a wedge. The broadening smears somewhat the minimum. As more detailed information we study the standard deviations along the ray $v t$. It grows as $\sqrt{t}$ (not displayed), unless one picks one of the peak velocities and the corresponding linear combination of currents. For the heat peak and current the growth as $t^{3/10}$ is confirmed and correspondingly the $t^{1/3}$ for the sound peak, see Fig.~\ref{fig:altern_mass_Jsite_time_std_deviation}.

\begin{figure}[!ht]
\centering
\subfloat[left sound peak]{
\includegraphics[width=0.31\textwidth]{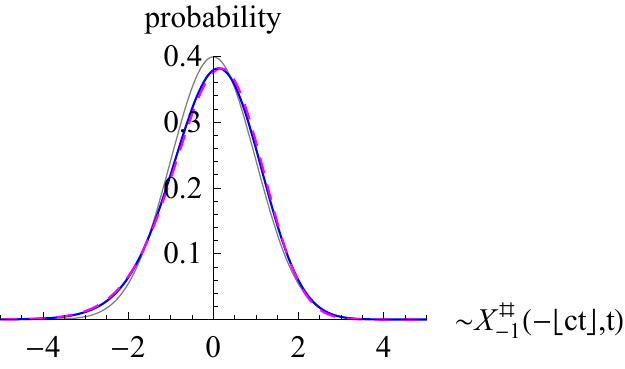}}
\hspace{0.01\textwidth}
\subfloat[heat peak]{
\includegraphics[width=0.31\textwidth]{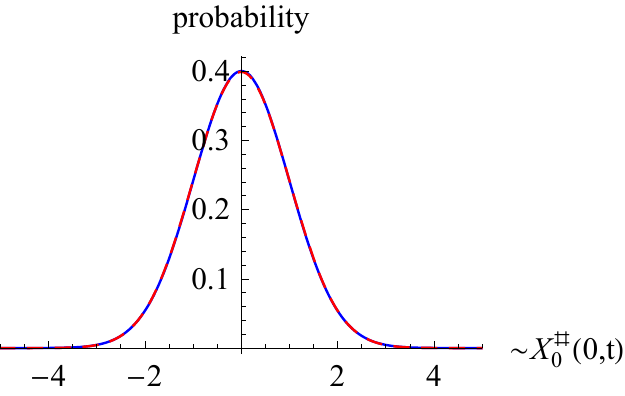}}
\hspace{0.01\textwidth}
\subfloat[right sound peak]{
\includegraphics[width=0.31\textwidth]{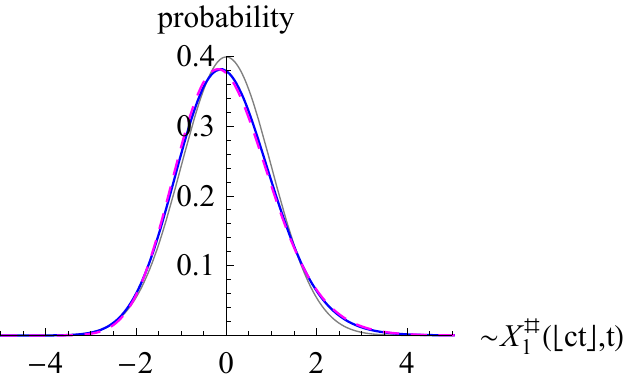}}\\
\subfloat[left sound peak]{
\includegraphics[width=0.31\textwidth]{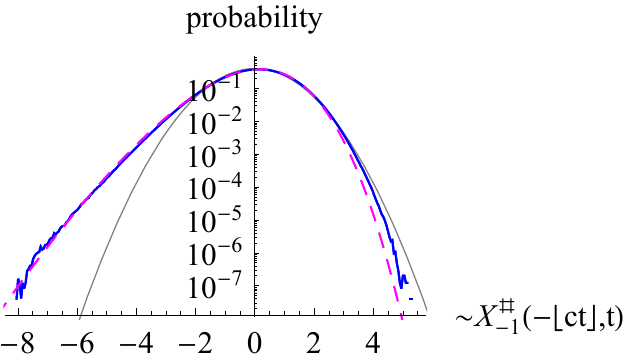}}
\hspace{0.01\textwidth}
\subfloat[heat peak]{
\includegraphics[width=0.31\textwidth]{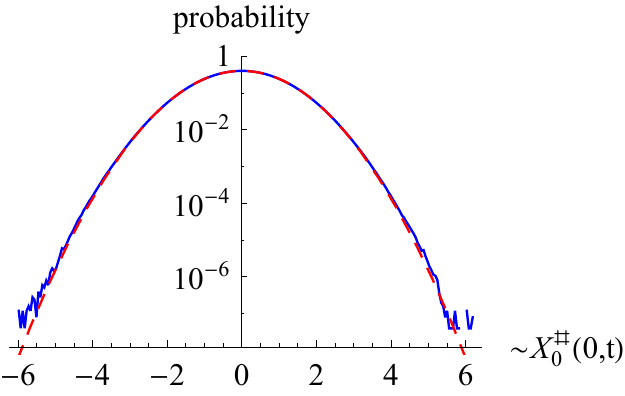}}
\hspace{0.01\textwidth}
\subfloat[right sound peak]{
\includegraphics[width=0.31\textwidth]{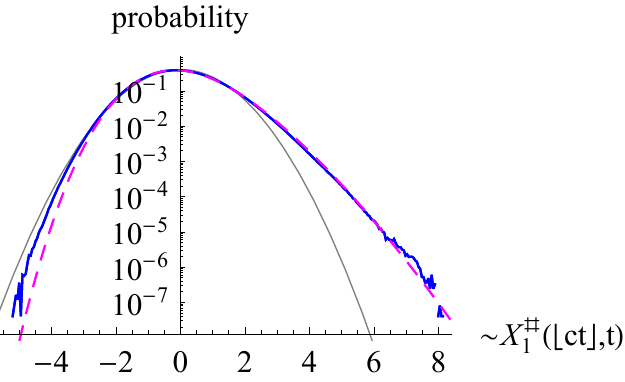}}
\caption{(Color online) Statistical distribution (blue) of the normal-mode currents integrated along the path \eqref{4.5} with end points $y = -ct,0,ct$. The theoretically predicted Baik-Rains distribution (dashed magenta curves) matches the sound peaks reasonably well. The gray thin curves show a normal distribution for comparison. The integrated current for the heat peak has a perfect normal distribution (red dashed in the center).}
\label{fig:altern_mass_Jsite_BaikRains}
\end{figure}
On the next level we study the full current statistics. In the non-exceptional cases we find indeed a normal distribution. Presumably the increments of these integrated currents are approximately independent and there is Brownian motion-like behavior in $t$. But also for the heat peak and its current we find a normal distribution, see Fig.~\ref{fig:altern_mass_Jsite_BaikRains}, although on the scale $t^{3/10}$. This suggests that, as a function of $t$, in the scaling limit one converges to Gaussian process with a yet to be determined covariance. The result for the sound peaks and currents is also displayed in Fig.~\ref{fig:altern_mass_Jsite_BaikRains}. One observes an asymmetric distribution. Its tail away from the origin fits precisely the tail of the Baik-Rains distribution. For the tail towards the origin the agreement is not equally accurate.

The theory also predicts the asymptotic distribution in case the end point is not exactly at $c t$, but at $c t + w\,t^{2/3}$ with $w\in\mathbb{R}$ \cite{PrSp04}. The limit distribution depends on $w$. Also joint distributions are known \cite{FerrariPeche2010}. For such fine details considerably more intense numerical efforts would have to be invested. For us it is truly amazing that a chain with Hamiltonian dynamics generates the well hidden Baik-Rains distribution.

\section{Summary and conclusions}\label{sec7}

We performed MD simulations for anharmonic chains with three distinct hard collision potentials and specifically studied the long-time decay of the total current-current correlations. This is a $3\times 3$ matrix, for which only the $2,3$ block is non-constant in time. In the many previous studies only the energy current correlations were investigated, a notable exception being the recent work by van Beijeren and Posch \cite{vBPo}, who in addition study the momentum current correlation for the chain with shoulder potential. We examine also the cross correlations which are antisymmetric in $t$ and have a rapid, presumably exponentially fast decay. As a standard the energy current correlation decays as $t^{-2/3}$, which has been noted in previous simulations. But the convergence can be slow and, depending on the model, other power laws could be guessed when relying only on intermediate time scales. As first noted in \cite{Spo}, based on nonlinear fluctuating hydrodynamics, the momentum current decays as $t^{-3/5}$, which is well confirmed in our simulations and as well in \cite{vBPo} for the shoulder potential. In fact the theory predicts also the non-universal coefficients. While not our primary goal, for the alternating masses with hard collisions the agreement is surprisingly good.

An interesting test of the theory results when certain coupling coefficients vanish and subleading contributions turn into leading ones. In principle the corrections could come from a more detailed analysis of the mode-coupling equations \cite{MeSp13}, but also from corrections to mode-coupling itself. One case is the off-diagonal $\Gamma_{23}$. The other one is the apparent $t^{-1}$ decay of the momentum current correlations for alternating masses with hard collisions. Of course the true asymptotics might turn out to be different. But the interesting point is that mode-coupling locates a model which convincingly has a decay faster than $t^{-2/3}$. This seems to be in contradiction to the exact second moment sum rule, since on the level of the correlator $S(j,t)$ the hard collision model does not show any significant difference from the other two models. But the second moment is sensitive to how the KPZ asymptotics of the sound peak is approached, and there is no general argument to rule out a decay different from $t^{-2/3}$.

From a more global perspective the identification of the Baik-Rains distribution is perhaps the most significant advance. While the full distribution will not be readily seen, Baik-Rains strongly supports the decoupling hypothesis. To say, unless the self-coupling vanishes, the fluctuations of the respective normal field component close to the sound peak are governed by the stochastic Burgers equation. For it Baik-Rains is a recently established theorem
\cite{BoCoFe14}.\medskip

\noindent\textbf{Acknowledgments}. HS thanks Harald Posch for instructive discussions at the Galileo Galilei Insitute, Firenze, and making the manuscript \cite{vBPo} available. CM thankfully acknowledges computing resources of the Leibniz-Rechenzentrum.

\appendix

\section{Speed of sound, $R$ matrix, and $G$ couplings}\label{secG}

Each model has distinct values for $c$, $R$ and $G$ at given pressure $P$ and inverse temperature $\beta$. For the monoatomic chain with shoulder potential and biatomic chain of hard-point particles, the numerical values for $c$, $R$ and $G$ can be found in the appendix of Ref.~\cite{MS14}. Considering the biatomic chain with square-well potential, we use the non-zero pressure $P = \frac{1}{8}$ (different from \cite{MS14}), and thus record $c$, $R$ and $G$ here. $G^{-1}$ is specified by the relation $G^{-1} = - (G^{1})^{\mathcal{T}}$, with ${}^\mathcal{T}$ denoting the transpose relative to the anti-diagonal. The entries are rounded to four digits for visual clarity.

Our parameters $P = \frac{1}{8}$ and $\beta = 2$ for the biatomic chain with square-well potential and alternating masses $m_0 = 1$, $m_1 = 3$ imply $ c = 1.7437$ and
\begin{equation}
R =
\begin{pmatrix}
-2.4407 & -0.7071 & 0.2028 \\
 0.3517 &  0      & 2.8139 \\
-2.4407 &  0.7071 & 0.2028 \\
\end{pmatrix}\,, \quad
R^{-1} =
\begin{pmatrix}
-0.2028 & 0.0292 & -0.2028 \\
-0.7071 & 0      &  0.7071 \\
 0.0253 & 0.3517 &  0.0253 \\
\end{pmatrix}\,,
\end{equation}
as well as
\begin{equation}
G^{1} =
\begin{pmatrix}
-0.0188 &  0.5969 & 0.1580 \\
 0.5969 & -0.1744 & 0.5969 \\
 0.1580 &  0.5969 & 0.1580 \\
\end{pmatrix}\,,
\quad
G^{0} =
\begin{pmatrix}
-1.2267 & 0 & 0      \\
 0      & 0 & 0      \\
 0      & 0 & 1.2267 \\
\end{pmatrix}. \\
\end{equation}


\end{document}